\documentclass[aps,prd,groupedaddress,superscriptaddress, twocolumn]{revtex4-2}

\pdfoutput=1

\usepackage{hyperref}
\usepackage{color}
\usepackage{graphicx}
\usepackage{array,booktabs}
\usepackage{float}
\usepackage{amsmath}
\usepackage{multirow} 
\usepackage{lineno}
\usepackage{comment}
\usepackage[normalem]{ulem}
\usepackage[caption=false]{subfig}

\begin{document}

\title{Calibration of the liquid argon ionization response to low energy electronic and nuclear recoils with DarkSide-50}

\newcommand{\Alberta}{Department of Physics, University of Alberta, Edmonton, AB T6G 2R3, Canada}
\newcommand{\APC}{APC, Universit\'e de Paris, CNRS, Astroparticule et Cosmologie, Paris F-75013, France}
\newcommand{\AQLNGS}{INFN Laboratori Nazionali del Gran Sasso, Assergi (AQ) 67100, Italy}
\newcommand{\AQGSSI}{Gran Sasso Science Institute, L'Aquila 67100, Italy}
\newcommand{\AQUni}{Università degli Studi dell'Aquila, L'Aquila 67100, Italy}
\newcommand{\AUM}{InstitutodeF\'õsica,Universidad Nacional Auto\'nomade M\'exico(UNAM), M\'exico 01000, Mexico}
\newcommand{\AstroCeNT}{AstroCeNT, Nicolaus Copernicus Astronomical Center, 00-614 Warsaw, Poland}
\newcommand{\Augustana}{Physics Department, Augustana University, Sioux Falls, SD 57197, USA}
\newcommand{\Belgorod}{Radiation Physics Laboratory, Belgorod National Research University, Belgorod 308007, Russia}
\newcommand{\BHSU}{School of Natural Sciences, Black Hills State University, Spearfish, SD 57799, USA}
\newcommand{\BINP}{Budker Institute of Nuclear Physics, Novosibirsk 630090, Russia}
\newcommand{\BNL}{Brookhaven National Laboratory, Upton, NY 11973, USA}
\newcommand{\BOINFN}{INFN Bologna, Bologna 40126, Italy}
\newcommand{\BOUniPHY}{Physics Department, Universit\`a degli Studi di Bologna, Bologna 40126, Italy}
\newcommand{\CAUniCHE}{Department of Mechanical, Chemical, and Materials Engineering, Universit\`a degli Studi, Cagliari 09042, Italy}
\newcommand{\CAUniPHY}{Physics Department, Universit\`a degli Studi di Cagliari, Cagliari 09042, Italy}
\newcommand{\CAINFN}{INFN Cagliari, Cagliari 09042, Italy}
\newcommand{\Carleton}{Department of Physics, Carleton University, Ottawa, ON K1S 5B6, Canada}
\newcommand{\Campinas}{Physics Institute, Universidade Estadual de Campinas, Campinas 13083, Brazil}
\newcommand{\CentroFermi}{Museo della fisica e Centro studi e Ricerche Enrico Fermi, Roma 00184, Italy}
\newcommand{\CIEMAT}{CIEMAT, Centro de Investigaciones Energ\'eticas, Medioambientales y Tecnol\'ogicas, Madrid 28040, Spain}
\newcommand{\Cluj}{National Institute for R\&D of Isotopic and Molecular Technologies, Cluj-Napoca, 400293, Romania}
\newcommand{\CPPM}{Centre de Physique des Particules de Marseille, Aix Marseille Univ, CNRS/IN2P3, CPPM, Marseille, France}
\newcommand{\CTLNS}{INFN Laboratori Nazionali del Sud, Catania 95123, Italy}
\newcommand{\ENUniCEE}{Engineering and Architecture Faculty, Universit\`a di Enna Kore, Enna 94100, Italy}
\newcommand{\ETHZ}{Institute for Particle Physics, ETH Z\"urich, Z\"urich 8093, Switzerland}
\newcommand{\FNAL}{Fermi National Accelerator Laboratory, Batavia, IL 60510, USA}
\newcommand{\FortLewis}{Department of Physics and Engineering, Fort Lewis College, Durango, CO 81301, USA}
\newcommand{\GEUni}{Physics Department, Universit\`a degli Studi di Genova, Genova 16146, Italy}
\newcommand{\GEINFN}{INFN Genova, Genova 16146, Italy}
\newcommand{\GlenEllyn}{Glen Ellyn, Illinois 60137, USA}
\newcommand{\Hawaii}{Department of Physics and Astronomy, University of Hawai'i, Honolulu, HI 96822, USA}
\newcommand{\Houston}{Department of Physics, University of Houston, Houston, TX 77204, USA}
\newcommand{\IHEP}{Institute of High Energy Physics, Beijing 100049, China}
\newcommand{\IPNO}{Institut de Physique Nucl\`eaire dÕOrsay, 91406, Orsay, France}
\newcommand{\INSTM}{Interuniversity Consortium for Science and Technology of Materials, Firenze 50121, Italy}
\newcommand{\IPHC}{IPHC, Universit\'e de Strasbourg, CNRS/IN2P3, Strasbourg 67037, France}
\newcommand{\JINR}{Joint Institute for Nuclear Research, Dubna 141980, Russia}
\newcommand{\Krakow}{M. Smoluchowski Institute of Physics, Jagiellonian University, 30-348 Krakow, Poland}
\newcommand{\Kurchatov}{National Research Centre Kurchatov Institute, Moscow 123182, Russia}
\newcommand{\Laurentian}{Department of Physics and Astronomy, Laurentian University, Sudbury, ON P3E 2C6, Canada}
\newcommand{\LNFINFN}{INFN Laboratori Nazionali di Frascati, Frascati 00044, Italy}
\newcommand{\LNHB}{Universit\'e Paris-Saclay, CEA, List, Laboratoire National Henri Becquerel (LNE-LNHB), F-91120 Palaiseau, France}
\newcommand{\Lodz}{Institute of Applied Radiation Chemistry, Lodz University of Technology, 93-590 Lodz, Poland}
\newcommand{\LPNHE}{LPNHE, CNRS/IN2P3, Sorbonne Universit\'e, Universit\'e Paris Diderot, Paris 75252, France}
\newcommand{\Manchester}{The University of Manchester, Manchester M13 9PL, United Kingdom}
\newcommand{\MEPhI}{National Research Nuclear University MEPhI, Moscow 115409, Russia}
\newcommand{\MIBIINFN}{INFN Milano Bicocca, Milano 20126, Italy}
\newcommand{\MIINFN}{INFN Milano, Milano 20133, Italy}
\newcommand{\MIPoliICA}{Civil and Environmental Engineering Department, Politecnico di Milano, Milano 20133, Italy}
\newcommand{\MIPoliCHE}{Chemistry, Materials and Chemical Engineering Department ``G.~Natta", Politecnico di Milano, Milano 20133, Italy}
\newcommand{\MIPoliEIB}{Electronics, Information, and Bioengineering Department, Politecnico di Milano, Milano 20133, Italy}
\newcommand{\MIPoliENE}{Energy Department, Politecnico di Milano, Milano 20133, Italy}
\newcommand{\MIUni}{Physics Department, Universit\`a degli Studi di Milano, Milano 20133, Italy}
\newcommand{\MSU}{Skobeltsyn Institute of Nuclear Physics, Lomonosov Moscow State University, Moscow 119234, Russia}
\newcommand{\NAINFN}{INFN Napoli, Napoli 80126, Italy}
\newcommand{\NAUniPHY}{Physics Department, Universit\`a degli Studi ``Federico II'' di Napoli, Napoli 80126, Italy}
\newcommand{\NAUniCHE}{Chemical, Materials, and Industrial Production Engineering Department, Universit\`a degli Studi ``Federico II'' di Napoli, Napoli 80126, Italy}
\newcommand{\NSU}{Novosibirsk State University, Novosibirsk 630090, Russia}
\newcommand{\OACINAF}{INAF Osservatorio Astronomico di Capodimonte, 80131 Napoli, Italy}
\newcommand{\Petersburg}{Saint Petersburg Nuclear Physics Institute, Gatchina 188350, Russia}
\newcommand{\PGUniCBB}{Chemistry, Biology and Biotechnology Department, Universit\`a degli Studi di Perugia, Perugia 06123, Italy}
\newcommand{\PGINFN}{INFN Perugia, Perugia 06123, Italy}
\newcommand{\PIINFN}{INFN Pisa, Pisa 56127, Italy}
\newcommand{\PIUniPHY}{Physics Department, Universit\`a degli Studi di Pisa, Pisa 56127, Italy}
\newcommand{\PNNL}{Pacific Northwest National Laboratory, Richland, WA 99352, USA}
\newcommand{\Princeton}{Physics Department, Princeton University, Princeton, NJ 08544, USA}
\newcommand{\Queens}{Department of Physics, Engineering Physics and Astronomy, QueenÕs University, Kingston, ON K7L 3N6, Canada}
\newcommand{\RHUL}{Department of Physics, Royal Holloway University of London, Egham TW20 0EX, UK}
\newcommand{\RMTreINFN}{INFN Roma Tre, Roma 00146, Italy}
\newcommand{\RMTreUni}{Mathematics and Physics Department, Universit\`a degli Studi Roma Tre, Roma 00146, Italy}
\newcommand{\RMUnoINFN}{INFN Sezione di Roma, Roma 00185, Italy}
\newcommand{\RMUnoUni}{Physics Department, Sapienza Universit\`a di Roma, Roma 00185, Italy}
\newcommand{\SAINFN}{INFN Salerno, Salerno 84084, Italy}
\newcommand{\SNOLabaddress}{SNOLAB, Lively, ON P3Y 1N2, Canada}
\newcommand{\SNOLAB}{SNOLAB, Lively, ON P3Y 1N2, Canada}
\newcommand{\SSUniCHP}{Chemistry and Pharmacy Department, Universit\`a degli Studi di Sassari, Sassari 07100, Italy}
\newcommand{\Sussex}{Physics and Astronomy, University of Sussex, Brighton BN1 9QH, UK}
\newcommand{\Temple}{Physics Department, Temple University, Philadelphia, PA 19122, USA}
\newcommand{\TNFBK}{Fondazione Bruno Kessler, Povo 38123, Italy}
\newcommand{\TNTIFPA}{Trento Institute for Fundamental Physics and Applications, Povo 38123, Italy}
\newcommand{\TNUni}{Physics Department, Universit\`a degli Studi di Trento, Povo 38123, Italy}
\newcommand{\TOINFN}{INFN Torino, Torino 10125, Italy}
\newcommand{\TOPoli}{Department of Electronics and Communications, Politecnico di Torino, Torino 10129, Italy}
\newcommand{\TOUni}{Physics Department, Universit\`a degli Studi di Torino, Torino 10125, Italy}
\newcommand{\TRIUMFaddress}{TRIUMF, 4004 Wesbrook Mall, Vancouver, British Columbia V6T2A3, Canada}
\newcommand{\TUM}{Physik Department, Technische Universit\"at M\"unchen, Munich 80333, Germany}
\newcommand{\UB}{Universiatat de Barcelona, Barcelona E-08028, Catalonia, Spain} 
\newcommand{\UCDavis}{Department of Physics, University of California, Davis, CA 95616, USA}
\newcommand{\UCLA}{Physics and Astronomy Department, University of California, Los Angeles, CA 90095, USA}
\newcommand{\UMass}{Amherst Center for Fundamental Interactions and Physics Department, University of Massachusetts, Amherst, MA 01003, USA}
\newcommand{\UOC}{Department of Chemistry, University of Crete, P.O. Box 2208, 71003 Heraklion, Crete, Greece}
\newcommand{\USP}{Instituto de F\'isica, Universidade de S\~ao Paulo, S\~ao Paulo 05508-090, Brazil}
\newcommand{\VTech}{Virginia Tech, Blacksburg, VA 24061, USA}

\author{P.~Agnes}\affiliation{\Houston}
\author{I.F.M.~Albuquerque}\affiliation{\USP}
\author{T.~Alexander}\affiliation{\PNNL}
\author{A.K.~Alton}\affiliation{\Augustana}
\author{M.~Ave}\affiliation{\USP}
\author{H.O.~Back}\affiliation{\PNNL}
\author{G.~Batignani}\affiliation{\PIINFN}\affiliation{\PIUniPHY}
\author{K.~Biery}\affiliation{\FNAL}
\author{V.~Bocci}\affiliation{\RMUnoINFN}
\author{W.M.~Bonivento}\affiliation{\CAINFN}
% ...

\collaboration{The DarkSide Collaboration}\noaffiliation





\begin{abstract}
DarkSide-50 has  demonstrated the high potential of dual-phase liquid argon time projection chambers  in exploring  interactions of   WIMPs in the  GeV/c$^2$ mass range. The technique, based on the detection of the ionization signal  amplified via electroluminescence in the gas phase,  allows to explore recoil energies  down to the sub-keV range.  We report here on the DarkSide-50 measurement of the   ionization yield of electronic recoils  down to $\sim$180~eV$_{er}$,  exploiting   $^{37}$Ar and $^{39}$Ar  decays, and extrapolated to a few ionization electrons with the Thomas-Imel box model. Moreover, we present a model-dependent determination of the ionization response to nuclear recoils down to $\sim$500~eV$_{nr}$, the lowest ever achieved in liquid argon, using \textit{in situ} neutron calibration sources and external datasets from neutron beam experiments.  
\end{abstract}


\maketitle

\section{Introduction}

 Dual-phase noble liquid time projection chambers (TPCs) have yielded, for more than a decade, ever increasing world leading  sensitivity for the search for Weakly Interacting Massive Particles (WIMPs) with mass greater than 10~GeV/c$^2$ \cite{Aprile:2018dbl, Cui:2017nnn, Akerib:2016vxi, Ajaj:2019imk, Agnes:2018fwg}. In recent years, the dual-phase technology has been extended to search for ``light'' dark matter candidates using  the  ionization component only, which allows to explore the sub-keV energy range.  Experiments like  XENON1T with a liquid xenon target \cite{Aprile:2019xxb, Aprile:2019jmx}  and DarkSide-50 with liquid argon (LAr)  \cite{Agnes:2018ves, Agnes:2018oej} have set the best limits on WIMP-nucleus  interactions for  M$_{\chi}$$>$1.8~GeV/c$^2$ (M$_{\chi}$$>$0.1~GeV/c$^2$ exploiting the Migdal effect), WIMP-electron scattering for M$_{\chi}$$>$30~MeV/c$^2$, and absorption of dark photons and axion-like particles for M$_{\chi}$$\gtrsim$0.2~keV/c$^2$. This was possible thanks to the intrinsic radiopurity, and high ionization yield and resolution, allowing for detection of single electrons, of noble liquids.

Unlike xenon, for which there exists a rich set of measurements that characterize its ionization response in the keV range (see, for instance. Refs \cite{Sorensen:2008ec, Goetzke:2016lfg, Akerib:2016mzi, Aprile:2017xxh, Aprile:2019dme}), the argon response is almost unexplored.  In spite of the smaller cross section of WIMP scattering on argon  compared to xenon, due to the lower atomic number,  interactions in argon produce more energetic recoils,  with a higher probability of being detected above the threshold.  The potential of dual-phase LAr TPC technology in direct dark matter search  can therefore be significantly enhanced through a better understanding of the ionization response, especially in the sub-keV region.   This represents the focus of this paper, through the measurement of the ionization response to electronic (ER) and nuclear (NR) recoils with DarkSide-50, using $\beta$-decay sources intrinsic to LAr, \textit{i.e.} $^{37}$Ar and $^{39}$Ar, and neutron sources located just outside  the   TPC,  specifically $^{241}$Am-$^{13}$C and $^{241}$Am-$^{11}$Be. In addition, external datasets from beam test experiments are included in the analysis to better constrain the NR ionization response. This approach, already adopted in the so-called DarkSide-50 \textit{low-mass} analyses published in 2018 \cite{Agnes:2018ves, Agnes:2018oej},  is presented in this paper in detail after being highly refined and improved.   

Such calibrations not only have the potential to improve current limits with DarkSide-50, but also provide a basis for future detectors specifically designed to search for  light dark matter candidates with LAr.

\section{The DarkSide-50 experiment}

The DarkSide-50 experiment is located in Hall C of the Gran Sasso National Laboratory (LNGS) in Italy. The TPC has a cylindrical  active LAr target  of $\sim$46~kg  The lateral walls are made of polytetrafluoroethylene (PTFE), surrounded by field shaping copper rings. Two arrays of 19 3-in diameter photomultiplier tubes (PMTs) are located on both ends behind the transparent anode and cathode respectively.  They observe light signals from both primary  scintillation (S1) and electroluminescence  (S2) from ionization electrons, extracted in the gas phase, after being  drifted with a 200 V/cm field in  liquid. The extraction efficiency is estimated $>$99.9\% at the extraction field of 2.8 kV/cm. The PTFE and the fused silica windows at the top and bottom of the cylinder are coated with tetraphenyl butadiene (TPB), a wavelength shifter, that absorbs the 128~nm  photons emission and re-emits softer photons with a peak wavelength at 420~nm.

The TPC was first operated, between 2013 and 2014, with atmospheric argon, and then until 2019, with low-radioactivity argon extracted from deep underground, naturally shielded against cosmogenic isotope production. For a more detailed description of the TPC, see ref. \cite{Agnes:2014bvk, Agnes:2015ftt}.  

The TPC is surrounded by a hermetic   neutron veto, a 4 m diameter stainless steel sphere filled with 30~t of liquid scintillator, loaded with trimethyl borate molecule (TMB) at 5\% mass fraction.    110 8-inch PMTs are mounted on the inner surface of the sphere to detect the scintillation light. Neutrons are mostly captured by  $^{10}$B present in TMB via the $^{10}$B(n, $\alpha$)$^7$Li$^*$ and $^{10}$B(n, $\alpha$)$^7$Li reactions,  with a mean capture time of $\sim$22~$\mu$s  \cite{Agnes:2015qyz},  thanks to the  large $^{10}$B cross section for thermal neutron. The first reaction, with 93.6\% branching ratio, is also responsible for the emission of a characteristic 478~keV $\gamma$ from the $^7$Li$^*$ de-excitation, efficiently detected thanks to the relatively high light yield of the scintillator ($\sim$530 pe/MeV). 

 The neutron veto is in turn immersed in 1000 t of ultra-pure water, instrumented as a Cherenkov veto against cosmic muons, and passive shield against external background. See ref. \cite{Agnes:2015qyz} for more details on the veto system.

\section{Detector response model}
\label{sec:response}

The DarkSide-50  ionization signals are affected  by instrumental effects, like the smearing induced by gas-phase electroluminescence  and by the PMT charge response.  The measured S2 yield ($g_2$), defined as the  mean number  of photoelectrons per ionization electron extracted in the gas pocket, and the associated relative resolution are   23$\pm$1~pe/e$^-$ and $\sim$27\%, respectively,  for events localized beneath the central PMT \cite{Agnes:2018ves}. An additional instrumental effect is related to the electron lifetime, \textit{i.e.}  the survival time of an electron to the capture from impurities in LAr along the drift. This was measured to be about 10~ms, almost 30 times longer than the maximum drift time in the TPC (376 $\mu$s \cite{Agnes:2018fwg}). The distortion of the ionization signal induced by the electron lifetime is therefore limited to a few percent. 

The dominant  instrumental effect  is the  dependence of the S2 response on the event radial position on the \textit{xy}-plane,  orthogonal to the electric field. The S2 response yields a factor of $\sim$4 difference between the centre and the edges of the TPC. This has been assessed using the 41.5~keV line from the $^{83m}$Kr gas source injected into the active mass, and applying the position reconstruction algorithm~\cite{Agnes:2017grb}, which provides a sub-cm-level resolution.  Such a distortion may be due either to a non-uniformity of the TPB thickness,  on the top fused-silica window, which would produce non-uniform conversion of VUV scintillation,  or to the sagging of the window itself, resulting in a varying thickness of the gas pocket,  and, in turn,  of the number of electron-luminescence photons produced.

S2 pulses can be  corrected using the radial-dependent efficiency measured with $^{83m}$Kr data. However, the energy range of interest for the \emph{low-mass} analysis extends down to $\sim$100~eV,  where  the reconstruction algorithm is  inefficient because of the low number of S2 photoelectrons. The analysis published in 2018 \cite{Agnes:2018ves}  overcame this issue by exploiting a  channel-based correction that defines  the  event position as that of the top-plane PMT (\emph{Ch$_{max}$}), which  observes the largest fraction of photoelectrons.  The position correction based on this definition is rather coarse, because it does not take the finite size (3-inches) of the DarkSide-50  PMTs into account.

To improve the accuracy of the measurement of the intrinsic LAr response to  ionization signals, all effects mentioned in this section have been incorporated into a Monte Carlo  simulation. This approach has been validated on a sample of $^{37}$Ar decays, naturally present in  LAr. This cosmogenic background decays  via a single electron capture transition, ground-state to ground-state, with a half-life of 35.01 (2) d \cite{TabRad_v7}.

The dominant $^{37}$Ar decay branches are from the electron capture on   K  (2.83 keV) and L1 (0.277 keV) atomic shells, with branching ratios of $\sim$90.4\% and $\sim$8.4\%, respectively,  determined with the \texttt{BetaShape} code \cite{Mougeot:2018whu, MOUGEOT2019108884} using the latest recommended Q-value of 813.87 (20) keV \cite{Wang_2021}.  The emitted cascades of electrons, X-rays, and UV photons   are  evaluated with the  \texttt{RELAX} software~\cite{relax}, based on the latest \texttt{EADL2017} library of atomic transition data  \cite{eadl2017}.   \texttt{RELAX} calculates atomic relaxation spectra of UV photons, X-rays and Auger electrons,  hereafter referred to as ``primaries'', due to bound-state-to-bound-state transitions for a single initial vacancy in the different sub-shells. From the atomic transition probabilities, a  deterministic step-by-step propagation of the vacancies was made up to the valence shell and to the neutralization. Atomic configuration has been accounted for, rejecting transitions that would require more electrons in a sub-shell than those actually present. It should be underlined that the only information available in the \texttt{EADL2017}  library is for a single atomic vacancy. In principle, each additional vacancy due to an Auger transition would require a complete recalculation of the atomic energies, wave functions and transition probabilities, which is not considered in this work.

The relaxation paths estimated by \texttt{RELAX}  are 5213 and 72 for the K and L1 shells, respectively.  Each contribution is determined per initial vacancy and  weighted by the corresponding capture probability. The average energies of Auger electrons, UV photons and X-rays grouped by shell, are quoted in Table   \ref{tab:37Ar}. We can assume that the primary particle is almost always an electron, either because  directly emitted or because extracted by  X-rays or UV photons via photoelectric effect.  The K- and L1-shells emit, on average 3.9 and 2.8 primaries, respectively, neglecting UV photons with energy not sufficient to photoionize an $^{40}$Ar atom. To evaluate the ionization yield, the mean number of primaries  will be subtracted from the number of detected electrons, as primaries are also drifted and extracted in the gas pocket  and contribute to the S2 signal.

\begin{table}[h]
\begin{tabular}{ l|ll|ll }
\hline
 & \multicolumn{2}{l}{K--shell EC} & \multicolumn{2}{l}{L1--shell EC}      \\
\hline
Branching Ratio & \multicolumn{2}{l|}{90.4\%} & \multicolumn{2}{l}{8.4\%}  \\
Total Released Energy & \multicolumn{2}{l|}{2829} & \multicolumn{2}{l}{277}  \\
Mean number of primaries\footnote{Excluding  undetectable via ionization} & \multicolumn{2}{l|}{3.9} & \multicolumn{2}{l}{2.8}   \\
\hline
  &  $\langle$N$\rangle$ & $\langle$E$\rangle$  &  $\langle$N$\rangle$ & $\langle$E$\rangle$ \\
\hline
K  Auger electrons  &    0.905    &        2414 & & \\
K X-rays                  &    0.095   &         2634\\
L  Auger electrons  &    1.77  &           179      &       0.9995    &      179\\
L X-rays                 &    8E-4      &       188         &    0.0005     &     207 \\
M  Auger electrons   &   0.35       &      51        &      0.96        &     51 \\
UV photons (E$>$16 eV)   &         0.77       &      25     &         0.86    &         25 \\
Undetectable via ionization & 3.26        &    13        &      2.10        &     13 \\
\hline
\end{tabular}
\caption{Average numbers ($\langle$N$\rangle$) and  energies ($\langle$E$\rangle$) of primaries (electrons, X-rays and UV photons) emitted in electron capture-induced cascades on $^{37}$Ar K and L1 shells as determined using \texttt{RELAX}~\cite{relax}  and  the \texttt{EADL2017} library \cite{eadl2017}. The capture probabilities were evaluated with  \texttt{BetaShape} \cite{Mougeot:2018whu, MOUGEOT2019108884}.  All energies are expressed in eV. }
\label{tab:37Ar}
\end{table}

The two lines from K and L1 electron capture are observed in the DarkSide-50 data  by subtracting the latest $\sim$500 days of the underground argon campaign dataset (from  2015 to  2018), where $^{37}$Ar is almost entirely decayed, from the first $\sim$100 days, normalizing the two samples by their livetimes.  The subtracted spectrum is fitted with simulated spectra, generated by independently varying the average number of  detected electrons ($N_e$) for each of the two  $^{37}$Ar lines. Events are simulated with a uniform spatial distribution in the TPC.  The intrinsic fluctuations induced by the number of emitted particles and by the LAr ionization response to ERs are  modelled 
with an  empirical fudge factor implemented as a Fano factor \cite{Fano:1947zz}. In the Monte Carlo, a fraction of electrons is  suppressed according to  the electron lifetime.  The surviving free electrons are extracted in the gas pocket  with 100\% efficiency, and converted into S2 photoelectrons with $g_2$ that varies depending on the \textit{xy}-position of the event, according to relative efficiency derived with the $^{83m}$Kr  calibration source.

\begin{figure}[t]
\centering
\includegraphics[width=1.0\columnwidth]{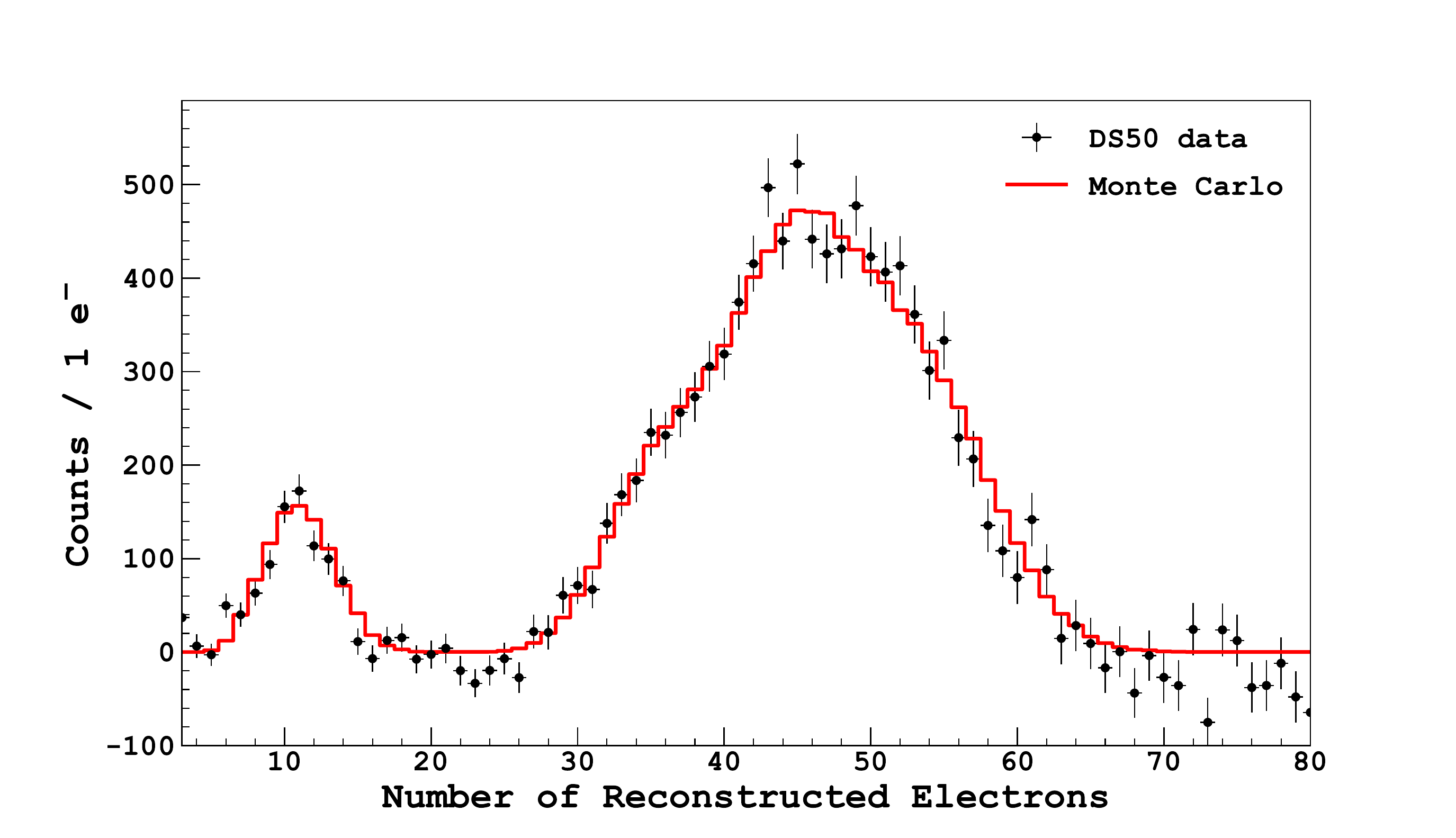}
\caption{Comparison between the best fitted simulated spectrum and $^{37}$Ar data as a function of the reconstructed number of  electrons. The free parameters in the model are the mean number of ionization electrons induced by the $^{37}$Ar L1- and K-shell electrons, and the Fano factor. }
\label{fig:37Ar}
\end{figure}

The fit is performed with  a $\chi^2$ analysis, where  the  free parameters are  the  numbers of   extracted electrons induced by the two $^{37}$Ar lines and the Fano factor. It is worth mentioning that no additional correction is applied to the resolution model. The shoulder-like structure observed at $\sim$35~$e^-$ in Figure \ref{fig:37Ar} originates from the non-uniform radial response of the detector and is well reproduced by the Monte Carlo starting from input parameters only.

The resulting numbers of  electrons   are 12.0$\pm$0.1(\emph{stat.})$\pm$0.5(\emph{syst.})  and 48.2$\pm$0.2(\emph{stat.})$\pm$2.1(\emph{syst.}) for the  $^{37}$Ar L1-- and K--shell, respectively.  The systematic errors are calculated by propagating the uncertainty on the $g_2$ value.  The best-fitted simulated spectrum is  shown in Figure \ref{fig:37Ar} together with data, as a function of the reconstructed number of electrons. The fit returns a reduced $\chi^2$ of 82.4 / 64.
 
 The ratio  between the measured amplitudes of the two lines, equal to 0.10$\pm$0.01, is in excellent agreement with $\sim$0.093, the expected value determined with  \texttt{BetaShape}, and with  measured values in literature  (0.103$\pm$0.003~\cite{PhysRev.120.2196},  0.102$\pm$0.004~\cite{doi:10.1080/14786436208212179}, 0.098$\pm$0.003~\cite{Totzek}).

 The fitted Fano factor, 0.10$\pm$0.03, is compatible with predictions on the LAr ionization fluctuation from the Shockley (0.107) and the Alkhazov (0.116) models   \cite{DOKE1976353}. This result suggests that  intrinsic fluctuations of the number of particles emitted in atomic cascades are negligible compared to the ionization fluctuation.

As already mentioned, in order to extract the number of ionization electrons from the two lines, the number of emitted primaries  in the atomic cascades must be subtracted  from the number of detected electrons. In the L1--shell cascade, the mean number of primaries is 2.8.  The 25 eV UV has high probability of extracting an electron by photoionization,  but  with insufficient energy to   induce additional ionization electrons.

The 51 eV M-shell electron is sufficiently energetic to ionize up to two atoms. The Thomas-Imel model, discussed in the following section, predicts a suppressed ion-electron recombination probability at such low ionization densities   \cite{PhysRevA.36.614}, a finding confirmed by numerical simulations \cite{Jaskolski:2011qja}. Yet the energy lost to excitations is unknown at such low energies. Therefore because of uncertainties on the ionization mechanism, we conservatively assume that the 51 eV line contributes with 1$\pm$1 ionization electrons to the L1-shell cascade.  In addition, we assume negligible interactions between the 51 eV and the L-shell 179 eV electron-ion clouds because of their low ionization density.  The resulting number of ionization electron at 179~eV is 8.2$\pm$1.3, where the uncertainty takes into account both statistical and systematic errors,  corresponding to a  ionization yield of  45.7$\pm$7.0~$e^-$/keV$_{er}$.

Unlike the L1--shell,  the superposition of $\sim$3 ion-electron clouds from K--shell electrons of 2414 eV (or 2634 eV X-rays)  and 179 eV (1.8 multiplicity) cannot be neglected. However, the lack of a model able to describe the complex event topology does not allow to estimate the overall recombination effect. For this reason, we do not include data from $^{37}$Ar K--shell in the analysis, discussed in the following section,  for the determination of  the ER ionization yield.

\section{Electronic recoil ionization yield}

The calibration of the ER energy scale relies on   $^{37}$Ar data, discussed in the previous section,  and on the cosmogenic $^{39}$Ar $\beta^-$-decay sample from the 2013-2014 atmospheric argon (AAr) campaign  \cite{Agnes:2014bvk}, acquired  with the same drift field of 200 V/cm. In AAr, $^{39}$Ar has a specific activity of $\sim$1~Bq/kg~\cite{Agnes:2014bvk}, and dominates the event rate. To suppress the  ``external" background from   radioactivity in  detector materials surrounding the active mass, events are selected within a central cylinder with 2~cm radius and 21.6~cm height, 16.8 cm far from the lateral walls and 7 cm from the top and  bottom of the TPC. The very narrow cut in radius selects events corresponding to the innermost area of the central PMT only, minimizing the non-uniformity of the detector response. To further  remove residual external contamination, events with more than one S2 pulse are rejected. These are  multiple scatter events, and not compatible with  the  topology of the $^{39}$Ar $\beta$-decay signature. 

The kinetic energy of each event is reconstructed exploiting the full  anti-correlation between the S1 and S2 signals, through  the so-called ``rotated" energy variable
\begin{equation}
E_{er} = w \, \left( \frac{S1}{g_1} + \frac{S2}{g_2}\right),
\label{eq:Eer}
\end{equation}

\noindent where $w$=19.5$\pm$1.0~eV \cite{Doke:2002oab} is the average energy required to produce a quantum (excitation or ionization)  and  $g_1$ the S1 collection efficiency (0.16$\pm$0.01 \cite{Agnes:2017grb}).  The number of  ionization electrons escaping the ion-electron recombination process, $N_{i.e.}$, is calculated for each event of the  $^{39}$Ar sample as 
\begin{equation}
N_{i.e.}= \frac{S_2}{g_2}  -1, 
\label{eq:Nie}
\end{equation}
\noindent which accounts for the subtraction of the primary electron from the $\beta$-decay. The mean value of the ER ionization yield, $Q_y^{ER}$, is estimated using  eqs.~\ref{eq:Eer} and \ref{eq:Nie} for each 0.2--keV$_{er}$ bin. A lower threshold of $E_{er}$$>$1.7~keV is applied  to guarantee 100\% efficiency in the identification of the S1 pulse.  The uncertainty is dominated by the systematics from the $g_2$ parameter.

The ionization yield per unit of ER energy from  $^{37}$Ar and $^{39}$Ar  data is defined as

\begin{equation}
Q_{y}^{ER} = \frac{N_{i.e.}}{E_{er}} = \frac{(1-r) N_i}{E_{er}},
\label{eq:qy_er}
\end{equation}

\noindent  where $N_i$ is  the number of produced electron-ion pairs. The electron recombination probability ($r$) is  predicted at low energies, $\mathcal{O}$(keV$_{er}$),  by the Thomas-Imel box model~\cite{PhysRevA.36.614},

\begin{equation}
1-r = \frac{1}{\gamma N_i} \ \textrm{ln}(1+\gamma N_i), 
\label{eq:thomasimel}
\end{equation}

\noindent where $\gamma$ is a  free parameter describing the recombination of the initial electron-ion pairs contained in a box and immersed in an electric field. This model has proven to work well in noble liquids for spatially short tracks  \cite{Sorensen:2011bd}, in the $\mathcal{O}$(keV$_{er}$) range for ERs. In this range, Eq. \ref{eq:qy_er} can be parametrized as
\begin{equation}
Q_{y}^{ER} =     \frac{1}{\gamma} \, \frac{\textrm{ln}(1+  \gamma \, \rho \, E_{er})}{E_{er}}  , 
\label{eq:qy_er2}
\end{equation}
\noindent with $\rho$=$N_i$/$E_{er}$ and $\gamma$ the free parameters of the model. $\gamma$ can be expressed  as $C_{box}/F$, where $F$ is the drift field (200~V/cm) and $C_{box}$ depends on the mean ionization electron  velocity and on the size of the ideal box containing the electron-ion cloud. In this parameterization, we  assume the approximation of a constant excitation-to-ionization ratio, which  implies that $N_i$ is proportional to the deposited energy  \cite{Agnes:2017grb}.  The impact of this approximation is discussed at the end of this section. 

\begin{figure}[t]
\centering
\includegraphics[width=1.0\columnwidth]{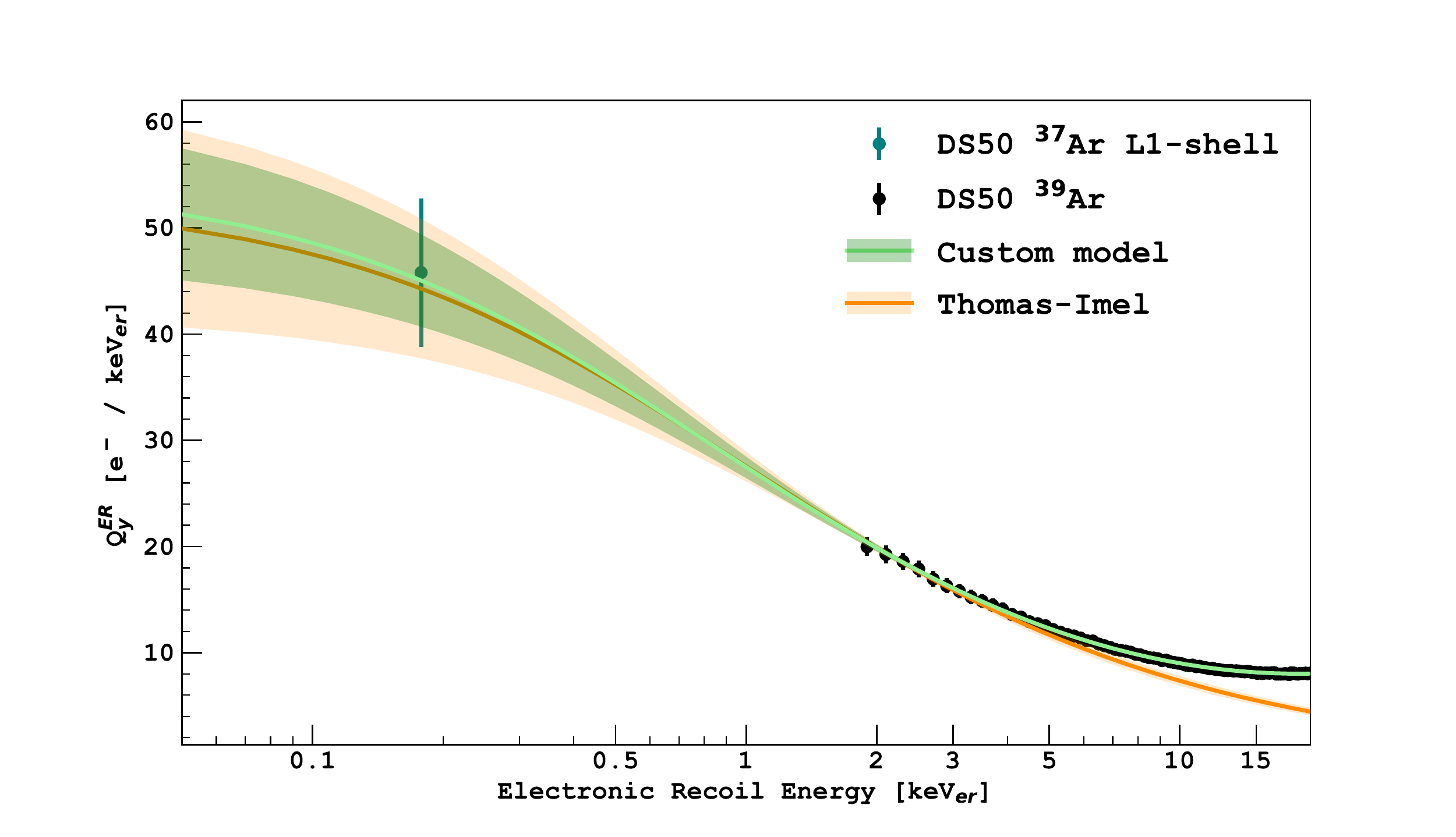}
\caption{Fit of the ER ionization yield, measured from AAr (black) and $^{37}$Ar (teal) data with a drift field of 200~V/cm, with the Thomas-Imel box model up to 3~keV$_{er}$ (eq.~\ref{eq:thomasimel}), and with the custom model from eq. \ref{eq:qy_er3}, which extends the Thomas-Imel model up to 20~keV$_{er}$. The model bands correspond to 1$\sigma$, also accounting for the correlation from the $g_2$ systematics, which dominates the experimental uncertainties.}
\label{fig:er_scale}
\end{figure}

$^{39}$Ar  data,  together with the  $^{37}$Ar  calibration line, are fitted with Eq. \ref{eq:qy_er2} up to 3~keV$_{er}$. As shown in Figure \ref{fig:er_scale}, the Thomas-Imel box model is in good agreement with $^{37}$Ar and $^{39}$Ar data points with E$_{er}$$<$3~keV$_{er}$.   

The extension of the model to higher energies through the empirical Doke-Birks parameterization \cite{Doke:1988dp},  in good agreement with data from the ARIS experiment above  $\sim$40 keV$_{er}$ \cite{Agnes:2018mvl}, is not compatible with $^{39}$Ar data  in the 3--20 keV$_{er}$ range. Instead, the agreement with $^{39}$Ar data is recovered in the whole data range, as shown in Figure \ref{fig:er_scale}, by adding a custom term to Eq. \ref{eq:qy_er2}, as follows

\begin{equation}
Q_{y}^{ER} =   \left(\frac{1}{\gamma} + p_0\,(E_{er}/ \textrm{keV$_{er}$} ) ^{p_1} \right)\, \frac{\textrm{ln}(1+ \gamma \, \rho \, E_{er})}{E_{er}},
\label{eq:qy_er3}
\end{equation}
\noindent with  two free parameters,   $p_0$ and $p_1$. 

The fit yields   $C_{box}$=9.2$\pm$0.9~V/cm, $\rho$=54.4$\pm$ 7.3~keV$_{er}^{-1}$, $p_0$=0.11$\pm$0.03, and $p_1$=1.71$\pm$0.08.  The    $C_{box}$ and $\rho$ parameters are compatible within 1$\,\sigma$ with 8.6$\pm$1.5~V/cm and 52.7$\pm$10.9~keV$_{er}^{-1}$, respectively,  obtained from the fit with the Thomas-Imel box model up to 3~keV$_{er}$.  It is worth highlighting that the extrapolation of the ER ionization yield below the  lowest measured energy  (179~eV$_{er}$ from the $^{37}$Ar L1-shell electron) is weakly dependent on the custom term introduced in Eq.~\ref{eq:qy_er3} as it is mainly driven by the Thomas-Imel box model.

To test the impact of the constant excitation-to-ionization ratio  assumption,  an energy-dependent parametrization  was introduced in Eq. \ref{eq:qy_er3},  so that 
\begin{equation}
\rho \rightarrow   \frac{\rho}{ 1 + \alpha(E_{er})},
\label{eq:rho}
\end{equation}

\noindent where 
\begin{equation}
\alpha(E_{er}) = \frac{0.21}{1 + e^{(E_{er} - b_1)/b_2}},
\end{equation}

\noindent is a sigmoid function tending to the  excitation-to-ionization ratio (N$_{ex}$/N{i}) of 0.21, as measured at  high energies \cite{Doke:1988dp}. The  fit of $^{37}$Ar and $^{39}$Ar data, assuming the change of variable as in eq. \ref{eq:rho}, requires two additional free parameters, and does not lead to  significant variations of $Q_{y}^{ER}$ with respect to the  constant $\rho$ approximation. Therefore, we conclude that  data are not sensitive to either assumption and opt to retain the formalism with constant N$_{ex}$/N{i} that represents the simplest model.


\section{Nuclear recoil ionization yield}

The NR ionization yield ($Q_{y}^{NR}$) is formalized in analogy with the ER one in eq. \ref{eq:qy_er}, 

\begin{equation}
Q_{y}^{NR} = \frac{N_{i.e.}}{E_{nr}} = \frac{(1-r) N_i}{E_{nr}}
\label{eq:qy}
\end{equation}

\noindent  where the electron recombination probability, $r$, is   described by the Thomas-Imel box model  in eq. \ref{eq:thomasimel}. 
%

Under the assumption that the excitation-to-ionization ratio is constant, $N_i$ can be expressed~\cite{Bezrukov:2010qa} as:

\begin{equation}
N_i = \beta \ \kappa(\epsilon) \ = \ \beta \ \frac{\epsilon \ s_e(\epsilon)}{s_n (\epsilon) + s_e (\epsilon)}
\end{equation}

\noindent  where $\beta$ is a normalization constant taken as the second free parameter of the model together with $C_{box}$.  The dimensionless parameter $\kappa$ represents the  energy lost in electronic excitations giving rise to ionization and scintillation signals. $s_e$ is the rate at which electrons  are excited by inelastic collisions while $s_n$ is the rate at which energy is transferred to recoiling  nuclei by elastic collisions. They all depend on the dimensionless parameter ($\epsilon$) defined as:

\begin{equation}
\epsilon = \frac{a}{2 e^2 Z^2} \ E_{nr}/\textrm{keV}\  \simeq \ 0.0135  \ E_{nr}/\textrm{keV}
\label{eq:eps}
\end{equation}

\noindent where the Thomas-Fermi screening length $a=0.626 \cdot a_0 \cdot Z^{-1/3}$~\cite{Bezrukov:2010qa} is used  with $a_0=\hbar/(\alpha m_e c) \simeq 0.529\cdot10^{5}$~fm. The stopping power, $s_e$, can be expressed as~\cite{Bezrukov:2010qa}:

\begin{equation}
\begin{aligned}
s_e(\epsilon) & = \frac{0.133 \ Z^{2/3}}{A^{1/2}} \ F(v/v_0) \ \sqrt{\epsilon}  \\ & \simeq \ 0.145 \ \ F(v/v_0) \  \sqrt{\epsilon}
\end{aligned}
\label{eq:s_e}
\end{equation}

\noindent where $F(v/v_0)$ is a correction factor dependent on the nuclear  ($v$) and Bohr ($v_0$ = $e^2$/$\hbar$) velocities.   With  no available  theoretical calculation backing either a suppression or an enhancement of the electronic stopping power, we assume $F(v/v_0)$=1. However, at the end of this section, we define an ad-hoc function as in Ref.  \cite{Bezrukov:2010qa} to test the sensitivity of the calibration data to a potential suppression of $s_e(\epsilon)$ at low energies.

The nuclear stopping power, $s_n$, is modelled by Ziegler et al. using an universal screening function~\cite{10.1007/978-3-642-68779-2_5}:

\begin{equation}
s_n(\epsilon) = \frac{\textrm{ln}(1+1.1383 f_Z  \, \epsilon)}{2[f_Z \, \epsilon+0.01321 (f_Z \, \epsilon)^{0.21226} + 0.19593(f_Z \, \epsilon)^{0.5}]}
\label{eq:ziegler}
\end{equation}

\noindent where $f_Z \simeq 0.953$ is a conversion factor for argon that accounts for the slightly different dimensionless energy definition used in Ref.~\cite{10.1007/978-3-642-68779-2_5} compared to Eq~(\ref{eq:eps}).

The NR response model is applied to  simulated events analogously to what done for  ERs in  Section \ref{sec:response}.  The only difference is in the intrinsic resolution of the NR ionization process. Following the procedure used in Ref. \cite{Agnes:2018ves}, we  considered two extreme models: one allowing for fluctuations in energy quenching, ionization yield, and recombination processes obtained with binomial  distributions,  and an other where the fluctuations in energy quenching are set to zero. The analysis described in the below cannot  distinguish between the two models, as the difference in the results is negligible. For this reason, in the following, we  only consider  the model without quenching fluctuations. 

The model from Eq. \ref{eq:qy} is constrained  by fitting DarkSide-50 calibration data, using  $^{241}$Am-$^{13}$C  and $^{241}$Am-$^{9}$Be neutron sources \cite{Agnes:2016yzy}, acquired during the underground argon (UAr) campaign, and  external datasets from the SCENE \cite{Cao:2014gns}, ARIS \cite{Agnes:2018mvl}, and Joshi \emph{et al.} \cite{Joshi:2014fna} experiments,    as described in the following.


\begin{figure}[h]
\centering
\includegraphics[width=1.0\columnwidth]{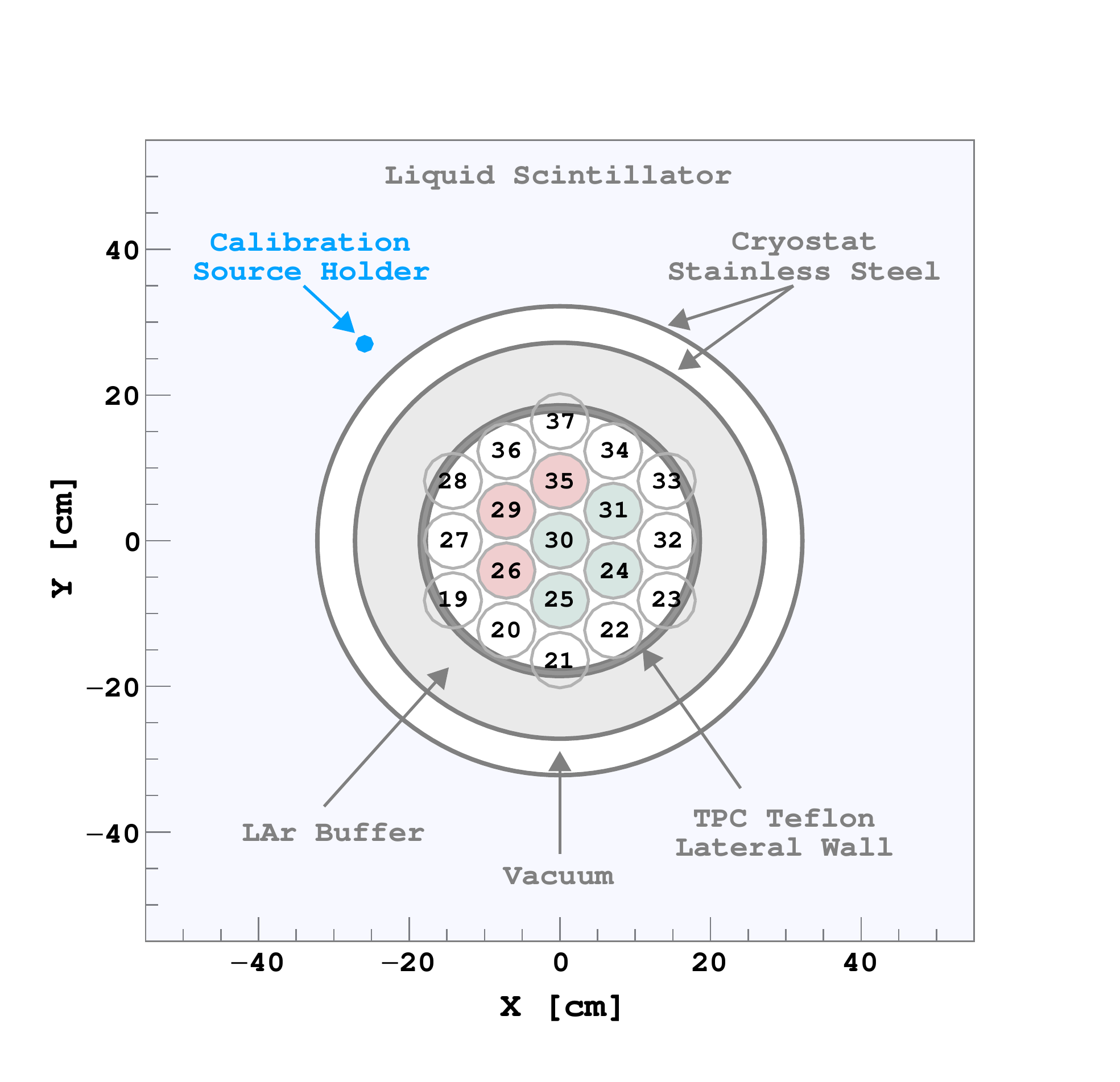}
\caption{Schematic top view of the detector. The source holder is located in the liquid scintillator veto. Source events have to cross the cryostat, the LAr buffer and the PTFE walls before reaching the active mass. Events whose position is associated  to channels in the outer ring (white) are not included in this analysis. Only events with maximum fraction of light observed by  PMTs highlighted in green  in the figure are included in the AmC analysis. The analysis of the  AmBe source is extended also to events selected  by PMTs highlighted in red.  }
\label{fig:tpc}
\end{figure}

\subsection{$^{241}$Am-$^{13}$C data selection}

The $^{241}$Am-$^{13}$C (from now on AmC) neutron source \cite{Liu:2015cra} is located outside the DarkSide-50 cryostat, in the liquid scintillator veto, as shown in Figure \ref{fig:tpc}. The source emits neutrons via ($\alpha$, n) on $^{13}$C, producing  $^{16}$O in the ground level  in the final state.  First or second excited states, accompanied by  $\gamma$ emission from  $^{16}$O$^*$  de-excitation, are   suppressed by a thin degrader, which  reduces the $\alpha$ energy below that needed to reach the lowest excited state of $^{16}$O.  

A 2 mm thick lead shielding   absorbs $^{241}$Am X-rays, resulting in a neutron source with very low correlation with $\gamma$ and X emission. However, to compensate for the low efficiency in neutron production, the $^{241}$Am activity is rather high ($\sim$3.6~MBq),  producing pile-up X-rays and $\gamma$s with a non-negligible probability of escaping the shielding and reaching the active volume of DarkSide-50, crossing the cryostat,  the LAr buffer surrounding the TPC, the  field-shaping copper rings, and the PTFE walls housing the active volume.

The  $^{241}$Am $\gamma$-ray with the highest branching ratio (35.9\%) has an energy  of 59.5~keV that falls in the regime dominated by the photoelectric effect  and   is fully absorbed in the LAr buffer and preceding materials. $^{241}$Am $\gamma$-rays with an energy $>$99~keV and branching ratio  $>$10$^{-9}$, as quoted in \cite{Firestone:391553}, are simulated with G4DS \cite{Agnes:2017grb}, the DarkSide-50 Monte Carlo, to derive the spectrum of AmC uncorrelated events. The detector response is applied as described in Section \ref{sec:response}.

\begin{figure}[]
\centering
\includegraphics[width=1.0\columnwidth]{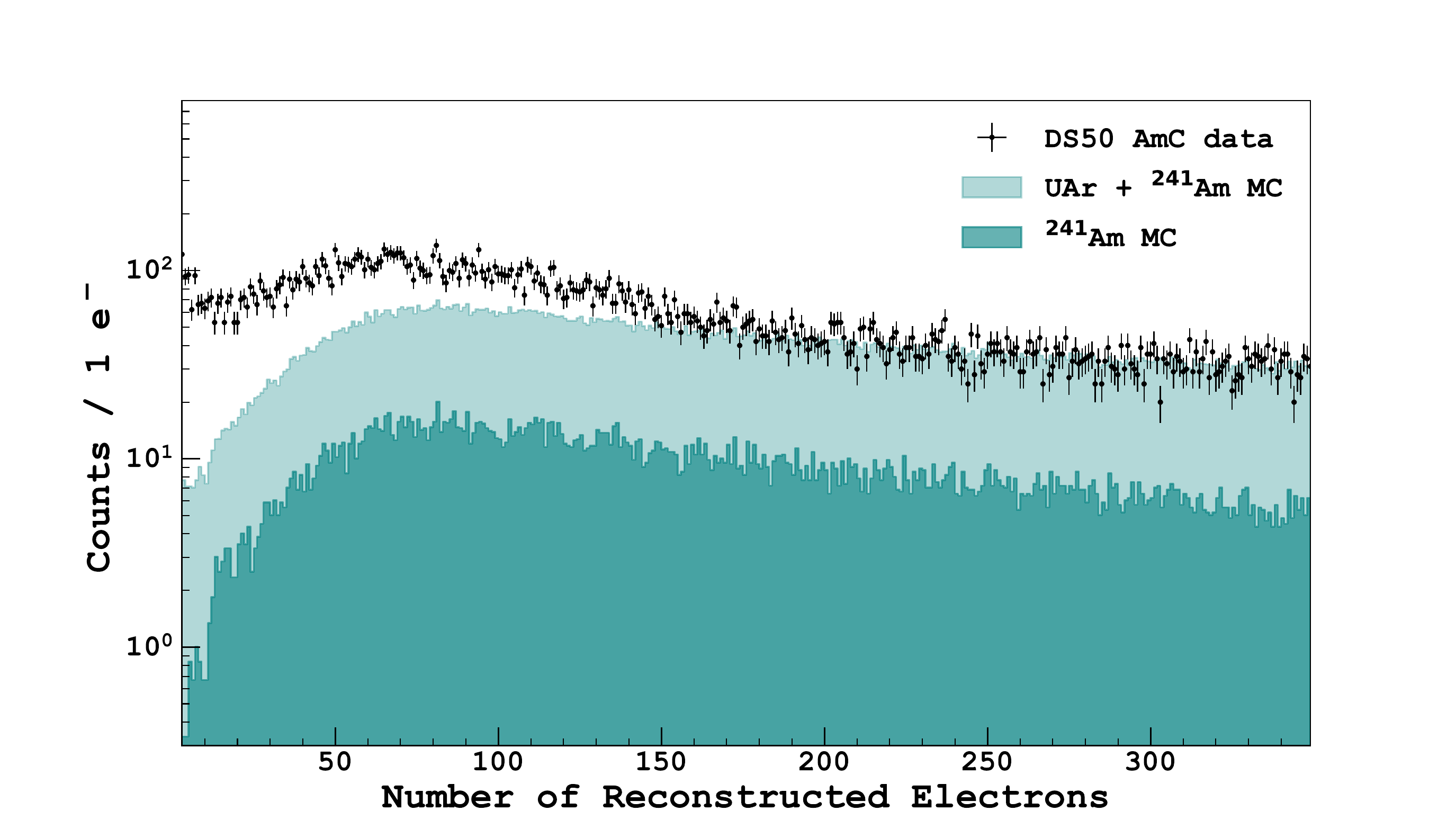}
\caption{Comparison between AmC spectrum (black) and contamination from TPC intrinsic events and $\gamma$s from $^{241}$Am, as described in the text. The low-energy excess is due nuclear recoils induced by neutrons.}
\label{fig:amc_spectrum}
\end{figure}

The AmC NR analysis is performed  on events with $Ch_{max}$ corresponding to channels 24, 25, 30, and 31, the 4 central PMTs less exposed to the source, as shown in Figure \ref{fig:tpc}, with  $^{241}$Am $\gamma$ contamination  minimized to 5.2\%, as estimated from a Monte Carlo simulation. Events selected by channels 26, 29, and 35 (see  Figure \ref{fig:tpc}) are excluded from the analysis because of the high rate of $^{241}$Am $\gamma$s.  Events not correlated with the source are subtracted using data from the UAr campaign, selected with the same {$Ch_{max}$ cuts and normalized by the livetime.

 Figure \ref{fig:amc_spectrum} shows AmC and UAr data spectra, the latter normalized by the livetime, and the simulated spectrum  of $\gamma$-events from $^{241}$Am reaching the TPC.  The uncertainty on the source position  affects the amplitude of the simulated spectrum, which is then normalized to the  UAr--subtracted AmC spectrum    in the  [250, 900]~$N_e$ range, where no NR event is expected. The excess of events from the source, shown in Figure \ref{fig:amc_spectrum}, with respect to the TPC intrinsic contamination and to simulated $^{241}$Am $\gamma$-rays,  is attributed to neutron scatterings. 



\subsection{$^{241}$Am-$^9$Be data selection}

\begin{figure}[t]
\centering
\includegraphics[width=1.0\columnwidth]{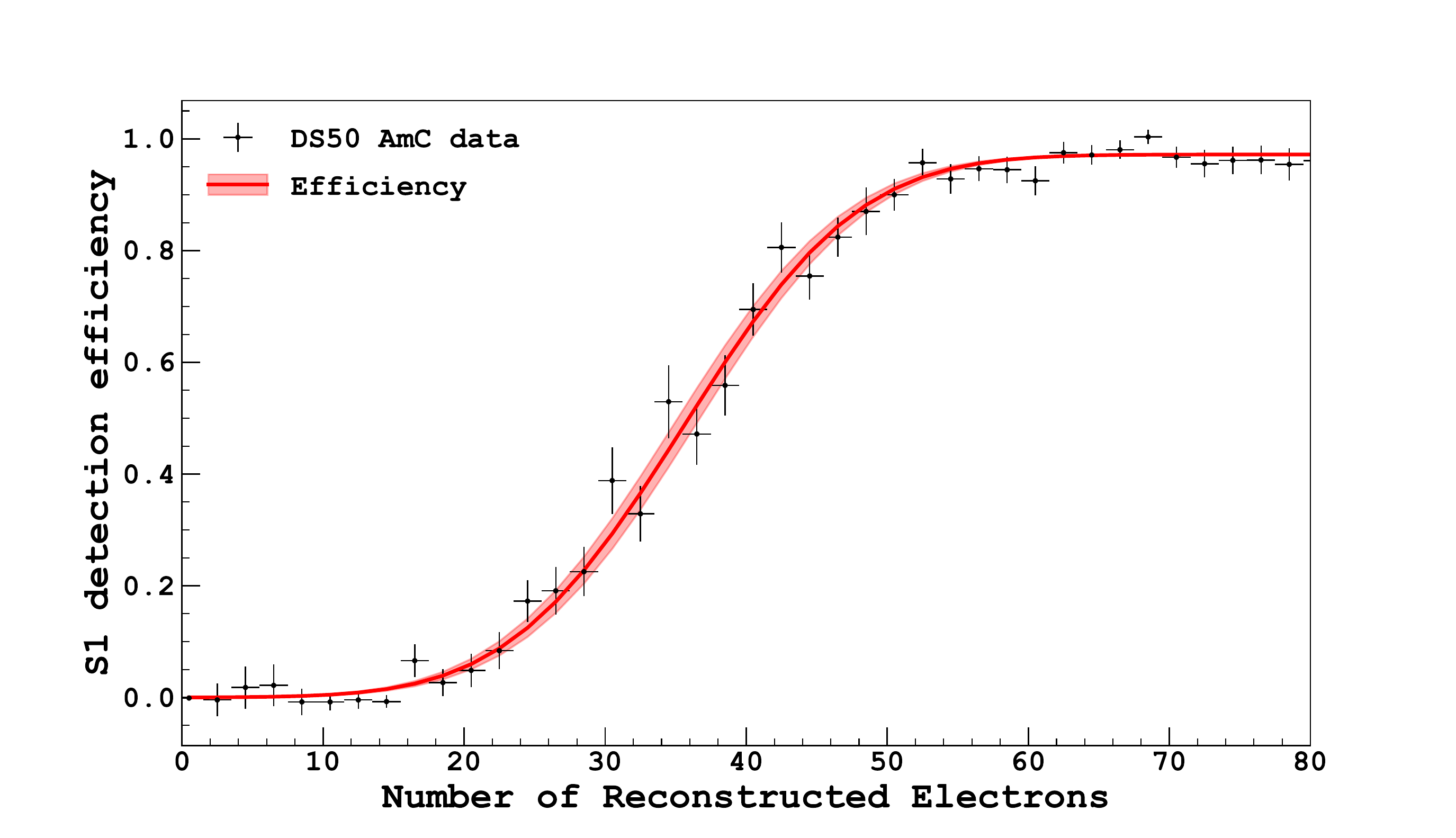}
\caption{Efficiency in detecting S1+S2 events (N$_{2p}$) with respect to S1+S2 and S2-only  (N$_{1p}$), as a function of the reconstructed number of electrons. Data are fitted with an error function (red line). The error band is propagated from the uncertainties of the fitted parameters.}
\label{fig:eff}
\end{figure}

The $^{241}$Am-$^9$Be source (from now on AmBe) emits  neutrons in association with the emission of, among others,  4.4~MeV $\gamma$s.  The prompt $\gamma$ signal is detected in the liquid scintillator veto \cite{Agnes:2015qyz}, where the source is deployed. Most neutrons, once scattered in the TPC LAr target, escape the TPC  given the low neutron capture cross section in $^{40}$Ar. Scattered neutrons that reach the neutron veto are mostly  captured by $^{10}$B with 22~$\mu$s mean time. 

\begin{figure}[t]
\centering
\includegraphics[width=1\columnwidth]{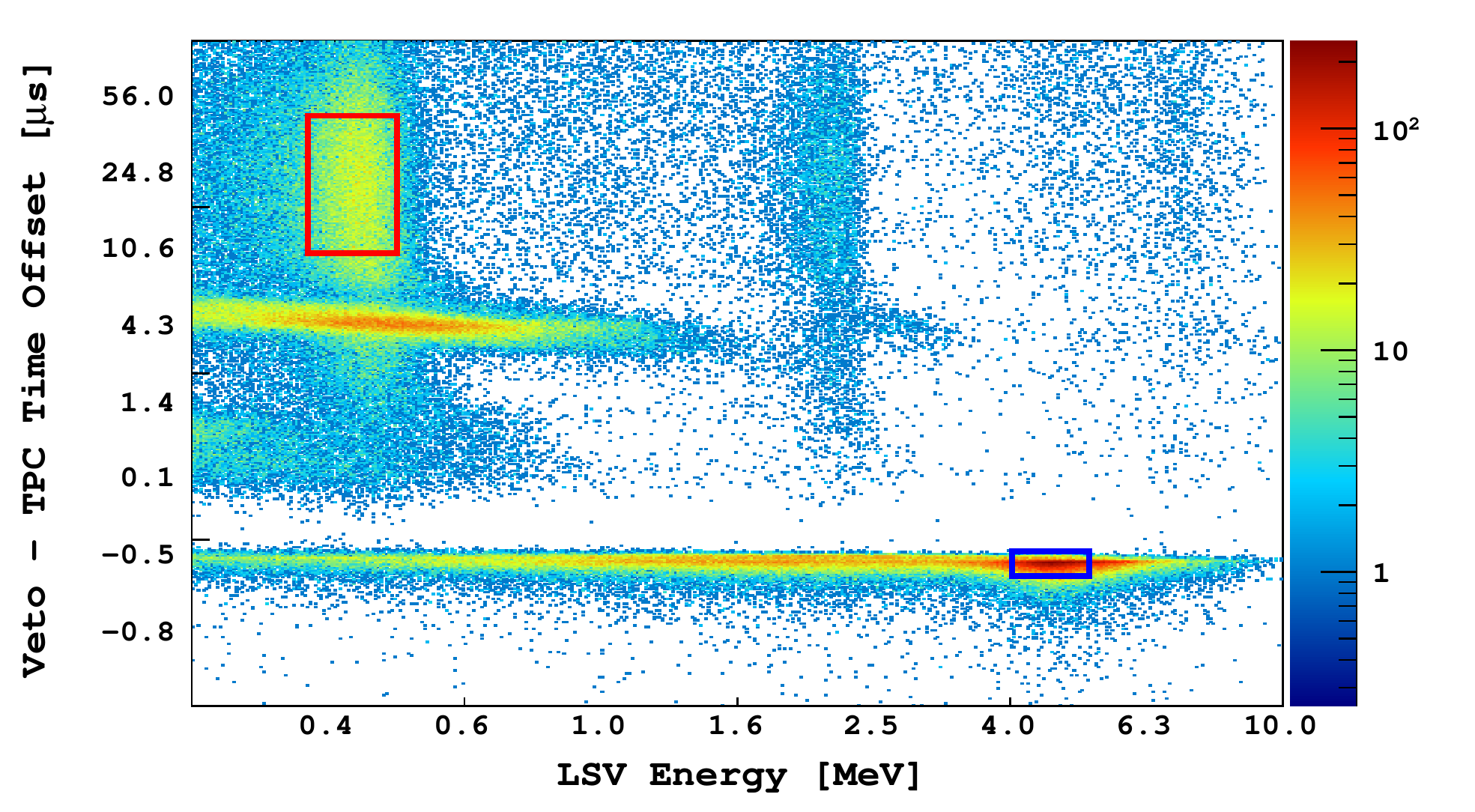}
\caption{Veto selection of prompt  (blue box)  and delayed (red box) events associated to AmBe neutron emission, as a function of liquid scintillator veto (LSV) visible energy and time difference between veto and TPC signals. The trigger time offset between the two detectors is about -550~ns. The horizontal feature at the 4.3~$\mu$s offset is due to  PMT afterpulses. }
\label{fig:ambe_selection}
\end{figure}

\begin{figure}[]
\centering
\includegraphics[width=1\columnwidth]{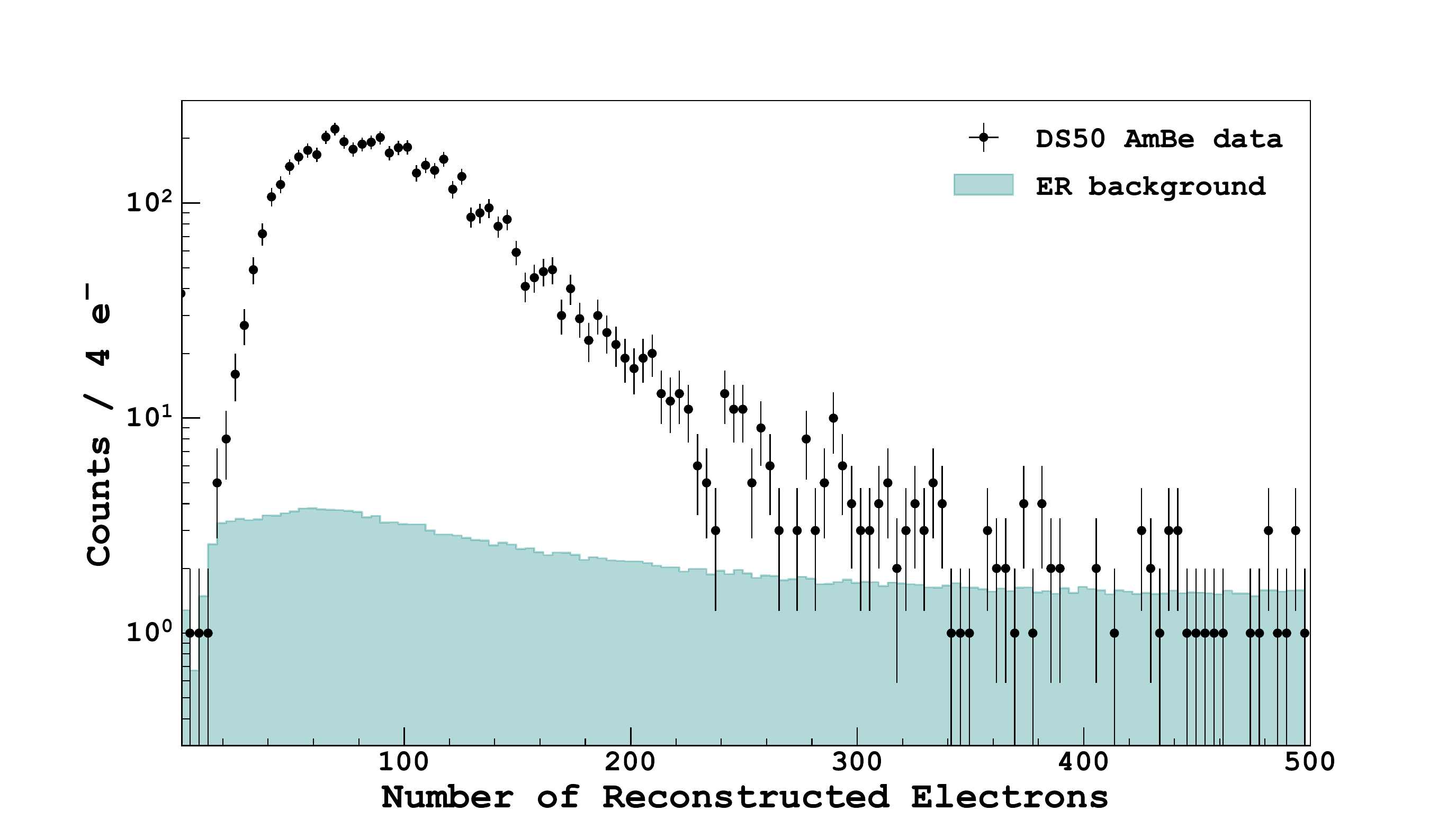}
\caption{Spectrum  of events selected in triple coincidence (black), compared with accidentals with $f_{90}$$<$0.4 from the UAr sample, normalized to events selected in the AmBe sample  with $N_e$ in the [200, 500] range.}
\label{fig:ambe_ne}
\end{figure}

NRs in the TPC are  selected with a three-fold coincidence by looking at the prompt 4.4~MeV  $\gamma$, followed by a single-scatter signal in the TPC within a few tens of nanoseconds, in turn followed by the delayed neutron capture in the veto. In order to apply this selection, events with both S1 and S2 pulses are required. The events with S2 signal only, in fact, are delayed by the drift time up to a maximum of 376~$\mu$s, much longer than the coincidence time with the prompt signal in the veto, which, in most cases, falls outside  the acquisition window. On the contrary, the S1 signal is very fast, and allows not to miss the coincidence with the  prompt signal in the veto.  The requirement of an S1 signal, however, implies an energy threshold, dictated by the  S1 trigger and pulse-finder efficiencies. The overall efficiency for NR events is defined as the ratio between  2-pulses events (S1+S2) and all events with either   S2-only or S1+S2. The efficiency is extracted from the AmC dataset, by subtracting the  ER contamination  from each of the two S2-only and S1+S2 samples. The resulting efficiency,  which from now on is named  S1 detection efficiency, is fitted with an error function, which takes into account Gaussian fluctuations expected in the S1 efficiency, as shown in Figure \ref{fig:eff}.

Prompt  signals in the veto are selected by requiring  an electron equivalent energy in the [4, 5]~MeV range, and a time difference with respect to the TPC between -600 and -500~ns, taking into account the   time offset of -550~ns between  veto and TPC. The delayed signal is selected in the [10, 45]~$\mu$s time window, and with event energy in the [370, 510]~keV range, a range properly tuned to account for quenching effects and $\gamma$ energy deposition in passive materials. The 10~$\mu$s low time threshold is applied to avoid veto PMT afterpulses. The selection cuts are shown in Figure \ref{fig:ambe_selection}. 

The AmBe  events selected in triple coincidence still suffer from a  residual  contamination due to accidental ERs. Such a contamination is account for  by looking in control region where no NR is expected. The control region is defined selecting events with   $N_e$   in the [200, 500]  range and with $f_{90}$$<$0.4, where  $f_{90}$ is the pulse shape discrimination estimator of DarkSide-50 \cite{Agnes:2014bvk},  less than 0.4, a region where no NR is expected. The  UAr spectrum is normalized with respect to the same selection criteria, and subtracted from the AmBe one. The contributions from  AmBe  and non-source events are shown in  Figure \ref{fig:ambe_ne}.


\subsection{External datasets}

A further constraint to the LAr response to the ionization signal is provided by ``external'' datasets, \textit{i.e.} measurements perfomed with small scale LAr detectors exposed to neutron beams. The SCENE   collaboration  \cite{Cao:2014gns} measured the ionization yield for 4 NR energies, between 16.9 and 57.3~keV, with $g_2$=3.1$\pm$0.3 pe/e$^-$. The drift field at 193 V/cm is very close to one used in DarkSide-50 (200 V/cm),  and the difference is assumed negligible in this analysis.  SCENE results are normalized  to the DarkSide-50 response by the ratio between the corresponding $g_2$'s. 

The ARIS collaboration   \cite{Agnes:2018mvl} characterized the LAr scintillation response at 200 V/cm for 8 NR energies, between 7.1 and 117.8 keV. ARIS S1 data at 200 V/cm are   rescaled to the DarkSide-50 response by the ratio between field-off S1 yields of DarkSide-50  (8.0$\pm$0.1 pe/keV) and ARIS (6.35$\pm$0.05 pe/keV).  This allows to associate the NR nominal energies from ARIS to the correspondent S1 at 200 V/cm in DarkSide-50. The final step is the conversion of the so-obtained S1 values  to S2, by looking at the S2/S1 ratio from NRs selected with the triple coincidence in the   AmBe dataset.  The correlation between S1 and S2 signals for the AmBe source was accounted for  by means of Monte Carlo simulations, which embeds the DarkSide-50  energy and optical response models \cite{Agnes:2017grb}. The resulting ionization yield is slightly lower at low energies than that published in 2018 \cite{Agnes:2018ves}. This difference is due to the improved modelling of the detector response.

Joshi \emph{et al.}  \cite{Joshi:2014fna} have measured the ionization yield of NRs at 6.7~keV using the endpoint of a spectrum induced by monochromatic 70~keV neutrons from a beam. The measurement at 240 V/cm, the closest field to the DarkSide-50 one, results in Q$_y$=3.6$\substack{+0.5 \\ -1.1}$~e$^-$/keV. However, after consulting with the authors, the data point is corrected for their single electron yield using the 2.82 keV K-shell capture $^{37}$Ar line from their experiment and DarkSide-50 as a cross-calibration point. The corrected value is then Q$_y$=6.0$\substack{+0.8\\ -1.8}$~e$^-$/keV. However, because of this correction, we preferred not to include  the measurement  by Joshi \emph{et al.}   in this analysis, but to quote it in the comparison with the final result.

\subsection{Global fit to data}

The energy Monte Carlo spectra for fitting both AmC and AmBe data samples are generated with G4DS, which accurately describes detector and source geometries, with a statistics of about $10^5$ neutron events, between 1 and 2 orders of magnitude higher than the data statistics. The detector response is applied with a toy Monte Carlo approach, as described earlier, by varying $\beta$ and $C_{box}$ parameters from the model with a fine scan. In addition, in order to take into account the inefficiency on S1 pulses, the efficiency curve shown in Figure \ref{fig:eff} is applied to AmBe simulated events as a function of $N_e$.

\begin{figure}[t]
\centering
\includegraphics[width=1.0\columnwidth]{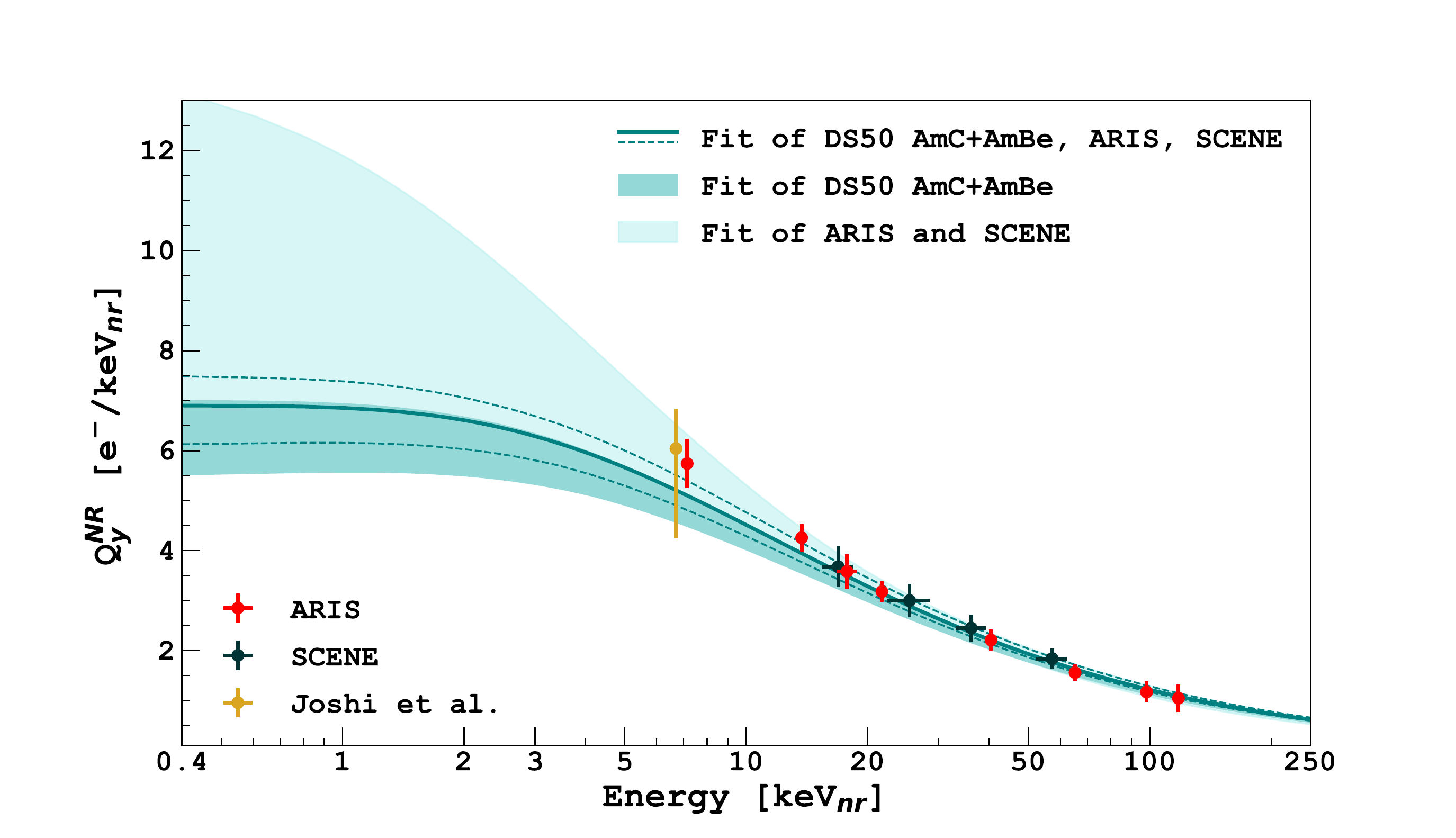}
\caption{Fit of the NR ionization yield at 200 V/cm from the combined fit of DarkSide-50 AmBe and AmC calibration data, together with  datasets from SCENE ~\cite{Cao:2014gns} and ARIS measurements~\cite{Agnes:2018mvl}, the latter combined with the DarkSide-50 ionization-to-scintillation ratio. The measured $Q_y^{NR}$ by  Joshi \emph{et al.}~\cite{Joshi:2014fna} at 6.7~keV$_{nr}$ is reported for comparison. The model bands correspond to 1~$\sigma$ uncertainty. }
\label{fig:fit_global}
\end{figure}

The AmC and AmBe datasets are simultaneously fitted with a $\chi^2$ analysis in the [3, 250]~$N_e$  range. The fit includes $g_2$ as an additional fit parameter constrained to the measured value within its uncertainty. To minimize the statistical fluctuations, the $\chi^2$ value is averaged over 800 toy simulations for each ($\beta$, $C_{box}$) parameter pair. The resulting $\chi^2$ map in the ($\beta$, $C_{box}$) parameter space is summed to the one from the simultaneous fit of the external datasets from SCENE and ARIS  presented in the previous section.

The resulting best parameters of such a global fit are $C_{box}$=$8.1\substack{+0.1 \\ -0.2}$~V/cm and  $\beta$=($6.8\substack{+0.1 \\ -0.3}$)$\times$10$^3$.   The corresponding model of the ionization yield is shown as a function of the nuclear recoil energy in Figure  \ref{fig:fit_global}. The errors on the model are from statistical uncertainties and  systematic error on $g_2$.   The $Q_y^{NR}$ extracted from external datasets only is in excellent agreement with the one from  AmC and AmBe calibrations, as shown in Figure \ref{fig:fit_global}, and the reduced $\chi^2$  from combined fit of all the datasets  is equal to 1.34 ($\chi^2$/NDF = 676/506).  Figure  \ref{fig:fit_am} shows the comparison between data and model, for both     AmC and AmBe, assuming best fitted parameters. The model is constrained at low energy by the  AmC calibration data, with a minimum energy value of $435\substack{+47 \\ -34}$~eV$_{nr}$, corresponding to 3 electrons. This is the lowest NR calibration threshold ever achieved in LAr.

\begin{figure}[t]
\centering
\includegraphics[width=1.0\columnwidth]{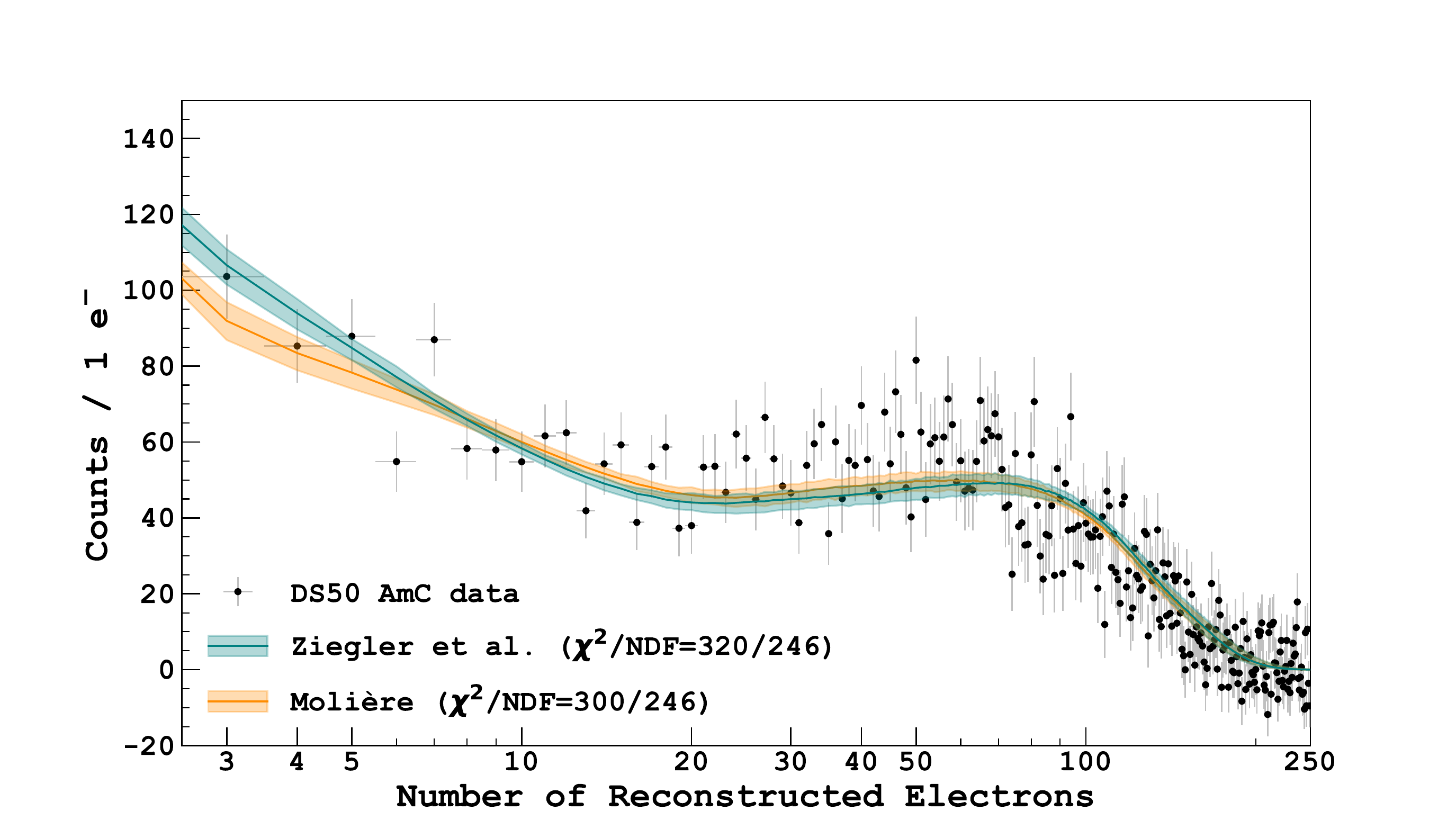}
\includegraphics[width=1.0\columnwidth]{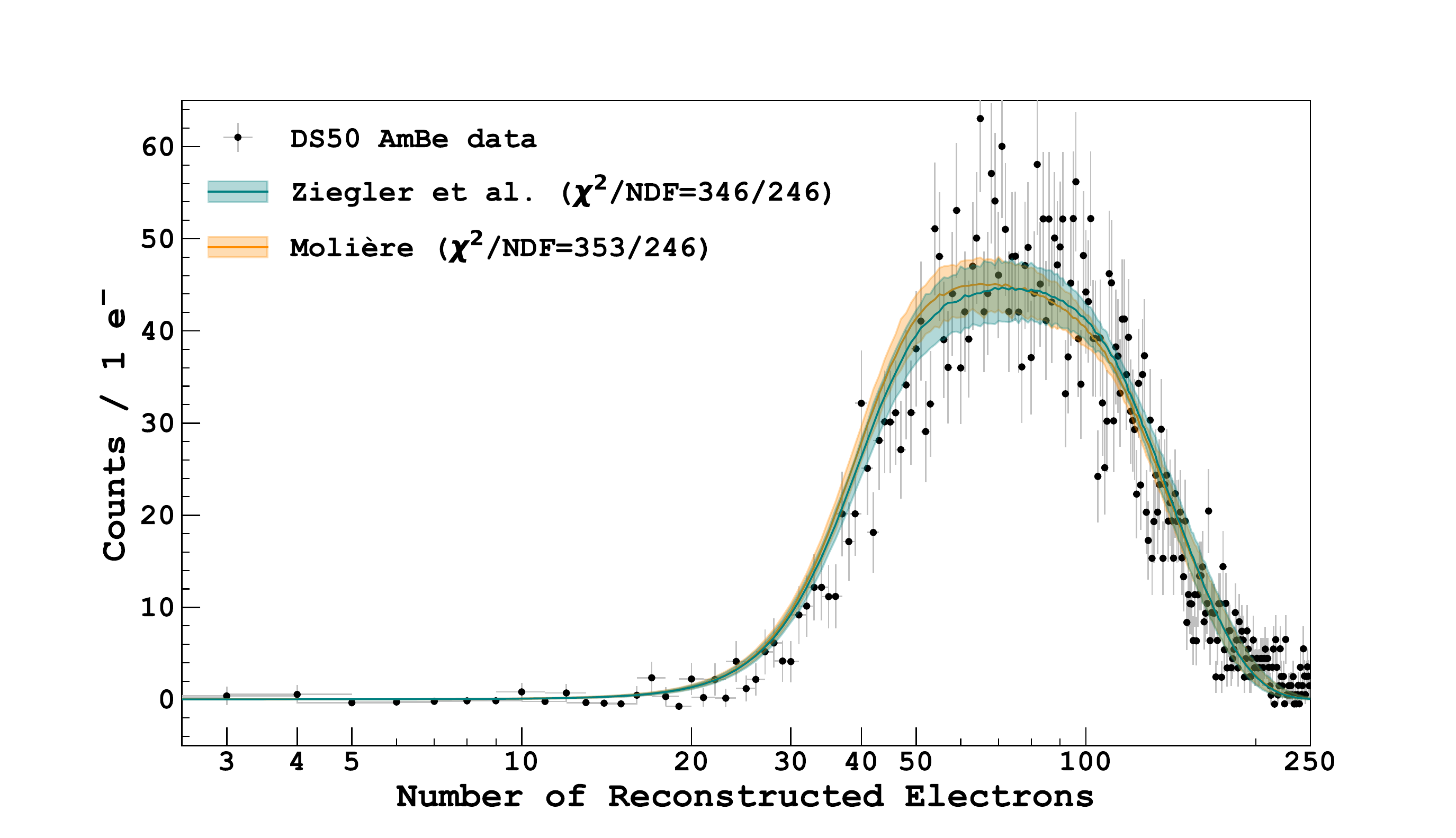}
\caption{Best fitted simulated $N_e$ spectra using Ziegler \textit{et al.} (blue) and Moli\`ere (orange) screening functions,  compared to AmC (top) and AmBe (bottom) data. The model bands correspond to 1$\sigma$ uncertainty. }
\label{fig:fit_am}
\end{figure}

Regarding potential systematics, we have investigated the impact of S1 detection efficiency on the result for 5 low thresholds, ranging from 0 to 40 $N_e$, applied to the AmBe dataset.  The results obtained by repeating the  fit for each threshold are in agreement within 1~$\sigma$ with the one without threshold, confirming that the S1 detection efficiency does not induce any appreciable systematics.

\subsection{Sensitivity to other theoretical models}

We have investigated the sensitivity of our data to different models of the nuclear stopping power as a function of energy stemming from different assumption on screening effects from atomic electrons. This is encapsulated in screening functions arising from different  models that can be written as  \cite{Bezrukov:2010qa}

\begin{equation}
f(\eta) = \frac{\lambda \, \eta^{1-2m}}{\left(1 + \left[  2 \, \lambda \, \eta^{2(1-m)} \right]^q  \right)^{1/q}},
\end{equation}

\noindent with each model characterised by a different set of parameters. The additional tested models, as suggested by Bezrukov \textit{et al.} \cite{Bezrukov:2010qa}, are Moli\`ere  ($m$ = 0.216, $q$ = 0.570, $\lambda$ = 2.37)  \cite{Moliere}, and Lenz-Jensen  ($m$ = 0.191, $q$ = 0.512, $\lambda$ = 2.92) \cite{lenz_uber_1932, Jensen}. Each of them leads to a different nuclear stopping power through

\begin{equation}
s_n(\epsilon) = \frac{1}{\epsilon} \int_0^{\epsilon} f(\eta)\,d\eta,
\end{equation}

\noindent with $\epsilon$ from Eq. \ref{eq:eps}. The Thomas-Fermi  screening function, also investigated by Bezrukov \textit{et al.} \cite{Bezrukov:2010qa},  is not included in this study as it is relevant when the projectile is a naked nucleus or an elementary particle, but not for partially ionized atoms. 

The  above-described analysis was performed  for the Moli\`ere  and Lenz-Jensen models and was found that both can fit successfully the data, with no statistically significant difference among them. The $Q_y^{NR}$'s obtained from the  fit using the different screening functions  are shown in Figure \ref{fig:fit_screenings}.   The comparison between model and data for  two   cases, Ziegler \textit{et al.}  and  Moli\`ere,   is shown in Figure \ref{fig:fit_am}.

The Ziegler \textit{et al.}  model is the one yielding the lowest  $Q_y^{NR}$ in the region of interest for WIMP analysis. Therefore the adoption of this model will result in the more conservative choice for the sensitivity to WIMPs  in future DarkSide searches.  

\begin{figure}[h]
\centering
\includegraphics[width=1.0\columnwidth]{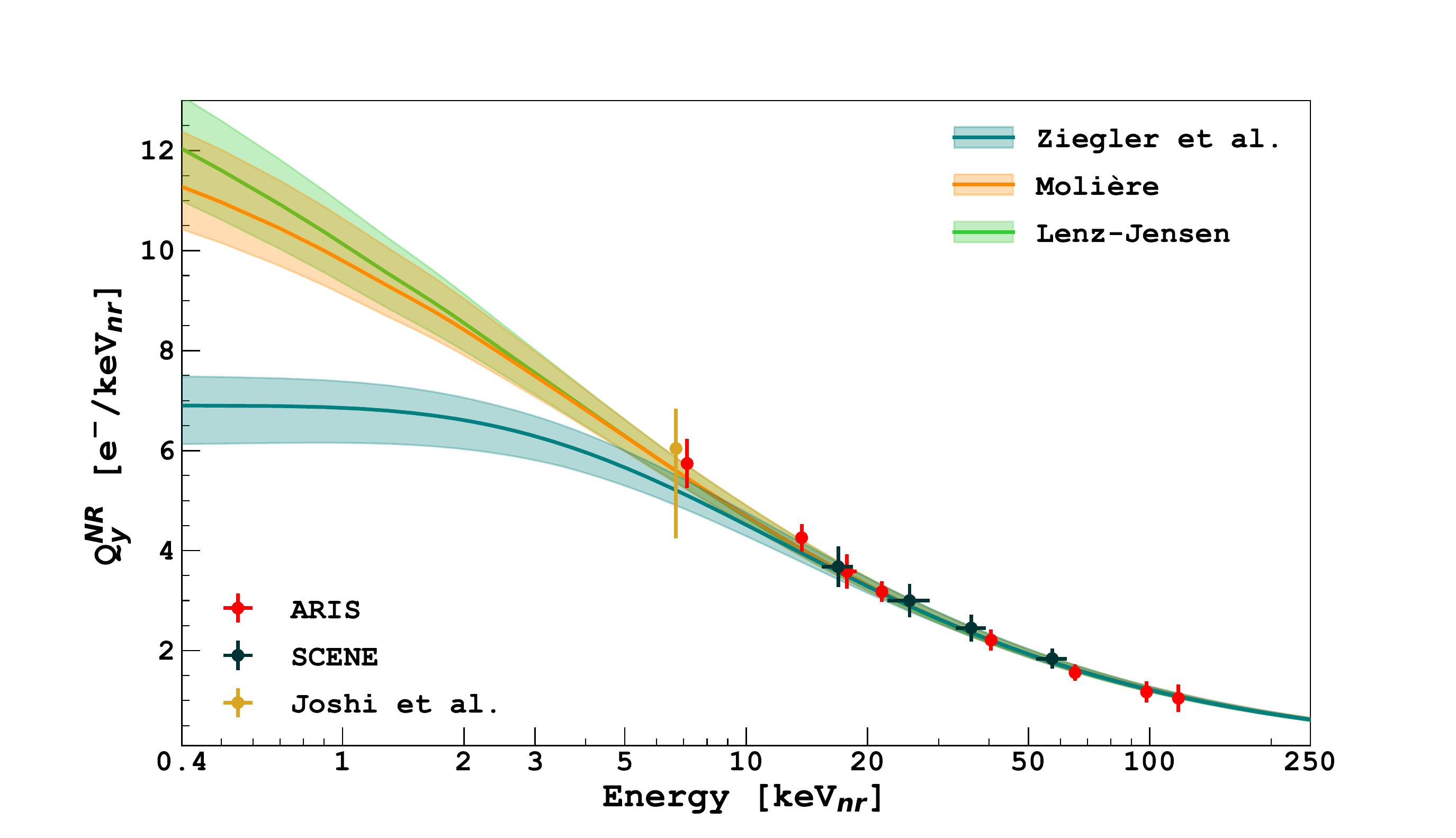}
\caption{Fit of DarkSide-50 AmBe and AmC  data, and ARIS and SCENE  datasets using Ziegler \textit{et al.}  (eq. \ref{eq:ziegler}) \cite{10.1007/978-3-642-68779-2_5}, Moli\`ere \cite{Moliere}, and Lenz-Jensen \cite{lenz_uber_1932, Jensen} screening functions. The model bands correspond to 1$\sigma$ uncertainty. }
\label{fig:fit_screenings}
\end{figure}

Following the same conservative approach, we also explored the impact of a low-energy $s_e$ suppression by introducing the  functional form of $F(v/v_0)$ in eq. \ref{eq:s_e}  as suggested in \cite{Bezrukov:2010qa}: 
\begin{equation}
F(v/v_0) = 1/2 \ (1 + \textrm{tanh}(50 \ \epsilon - z)),
\label{eq:F}
\end{equation}
\noindent where $F(v/v_0)$$\rightarrow$1 for $z\rightarrow -\infty$. Setting $z$ = 0.25 enables to reproduce the attempt in  \cite{PhysRevA.51.3058} to include Coulomb effects in the calculation of the electronic stopping power. Such a suppression, demonstrated to be not compatible with existing LXe datasets  \cite{Bezrukov:2010qa},   affects the energy range in LAr below $\sim$3~keV$_{nr}$, as shown in Figure \ref{fig:fit_sup}.  Through a null hypothesis test, where the null hypothesis corresponds to $F(v/v_0)$=1, we verified that  low energy AmC data   have the power to constrain suppression effect to be within $\sim$18\% (2$\, \sigma$) at 1 keV$_{nr}$ below the nominal charge yield derived from the universal function. This test allows to  exclude $z>$ 0.04 or $Q_y^{NR}$(1~keV$_{nr}$) $<$ 5.6 $e^-$ / keV$_{nr}$ at 95\% C.L..


\begin{figure}[h]
\centering
\includegraphics[width=1.0\columnwidth]{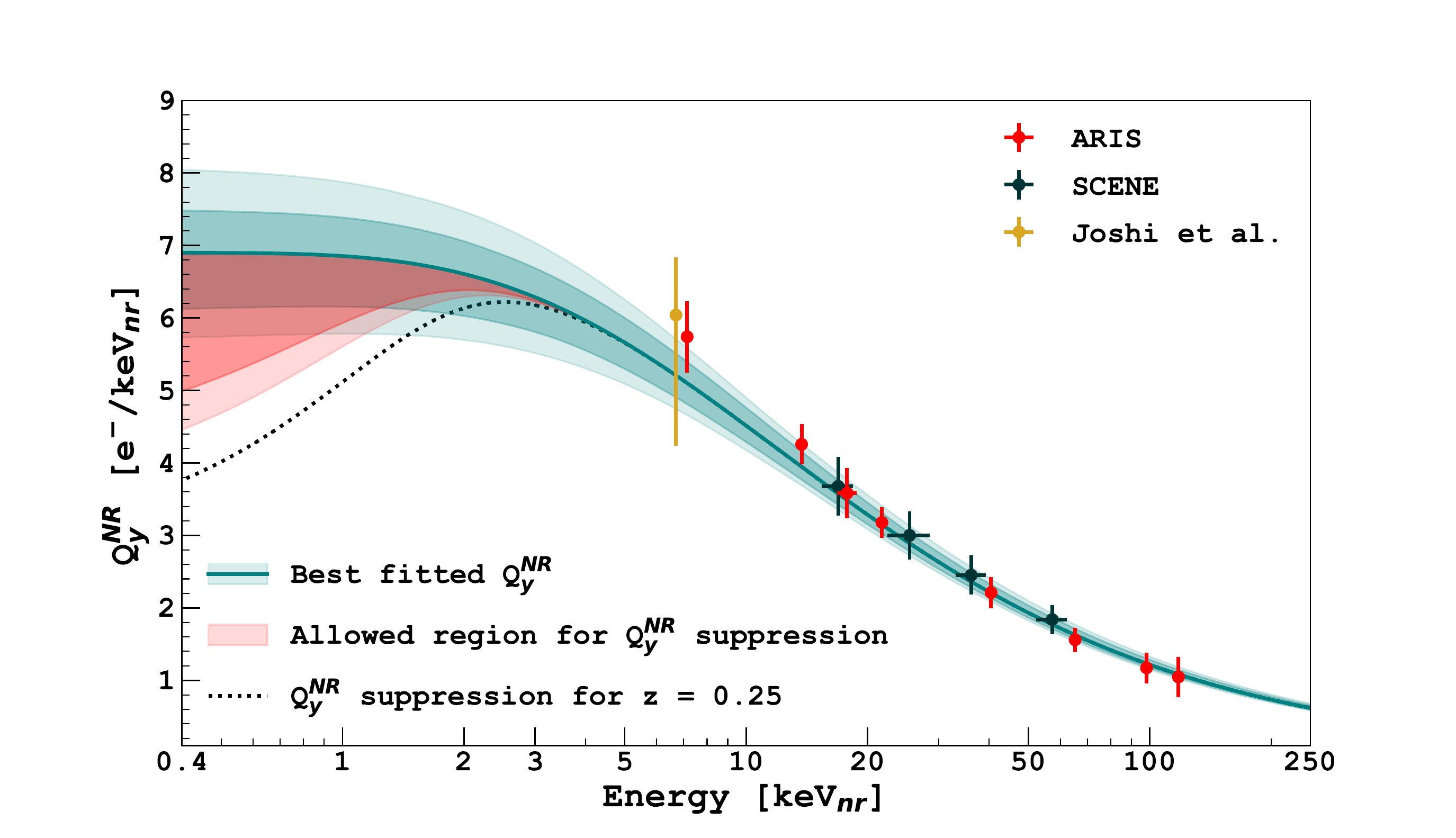}
\caption{Allowed region (red) for low energy $Q_y^{NR}$ suppression tested on the AmC dataset, and  best fitted  $Q_y^{NR}$  (blue), using the Ziegler \textit{et al.}  screening function. Dark (light) shaded area corresponds to 1$\sigma$ (2$\sigma$)  uncertainty. The dashed line represents the  $Q_y^{NR}$ suppression tested in \cite{Bezrukov:2010qa} assuming eq. \ref{eq:F} with $z$=0.25.  }
\label{fig:fit_sup}
\end{figure}

\section{Conclusions}

In this work, we presented the calibration of the LAr ionization response  to electronic and nuclear recoils in the keV region at 200 V/cm with DarkSide-50. The electronic recoil one is measured down  to 179~eV$_{er}$,    the energy of the L1-shell Auger electron  from $^{37}$Ar, and extrapolated down to few tens of eV by fitting data with the Thomas-Imel box model. The nuclear recoil ionization response is measured with a low threshold of  $\sim$500~eV$_{nr}$, corresponding to 3 ionization electrons, the lowest ever performed in LAr.   The measured ER and NR ionization yields will impact  direct dark matter searches  with LAr, extending the observation window to  low-mass  candidates, like  Weakly Interacting Massive Particles of few GeV/c$^2$ mass, axion-like particles, dark photons and sterile neutrinos, and to neutrino bursts from  core-collapse supernovae \cite{Agnes:2020pbw}. Dedicated  campaigns of measurement with setups exposed to neutron beams are highly desirable in the future to improve and better constrain response models at the keV scale.

\begin{acknowledgements}
We are grateful to Francesc Salvat and Pascal Lablanquie for their comments and suggestions.  The DarkSide Collaboration offers its profound gratitude to the LNGS and its staff for their invaluable technical and logistical support. We also thank the Fermilab Particle Physics, Scientific, and Core Computing Divisions. Construction and operation of the DarkSide-50 detector was supported by the U.S. National Science Foundation (NSF) (Grants No. PHY-0919363, No. PHY-1004072, No. PHY-1004054, No. PHY-1242585, No. PHY-1314483, No. PHY-1314501, No. PHY-1314507, No. PHY-1352795, No. PHY-1622415, and associated collaborative grants No. PHY-1211308 and No. PHY-1455351), the Italian Istituto Nazionale di Fisica Nucleare, the U.S. Department of Energy (Contracts No. DE-FG02-91ER40671, No. DEAC02-07CH11359, and No. DE-AC05-76RL01830), the Polish NCN (Grant No. UMO-2014/15/B/ST2/02561) and the Foundation for Polish Science (Grant No. Team2016-2/17). We also acknowledge financial support from the French Institut National de Physique Nucl\'eaire et de Physique des Particules (IN2P3),   the  IN2P3-COPIN consortium (Grant No. 20-152),  and the UnivEarthS LabEx program (Grants No. ANR-10-LABX-0023 and No. ANR-18-IDEX-0001),  from the São Paulo Research Foundation (FAPESP) (Grant No. 2016/09084-0),  from the Interdisciplinary Scientific and Educational School of Moscow University ``Fundamental and Applied Space Research'',  from the Program of the Ministry of Education and Science of the  Russian  Federation  for  higher  education  establishments,  project No. FZWG-2020-0032 (2019-1569), and from IRAP AstroCeNT funded by FNP from ERDF. Isotopes used in this research were supplied by the United States Department of Energy Office of Science by the Isotope Program in the Office of Nuclear Physics.
 \end{acknowledgements}

\bibliography{biblio}
\end{document}